\documentclass[12pt]{article}
\usepackage{a4}
\usepackage{amsfonts}

\newcommand{\be}{\begin{equation}}
\newcommand{\ee}{\end{equation}}

\newcommand{\ov}{\overline}
\newcommand{\re}{{\rm e}}

\newcommand{\vp}{\wedge}

\newcommand{\rD}{{\nabla}}
\newcommand{\CP}{{\mathbb C\, \mathbb P}}

\newcommand{\C}{{\mathbb C}}

\newcommand{\di}{{\rm dim}\,}

\newcommand{\tr}{{\rm tr}}
\newcommand{\ri}{{\rm i}}
\newcommand{\rd}{{\rm d}}
\newcommand{\al}{\alpha}

\newcommand{\p}{\partial}
\newcommand{\str}{\rule{0ex}{1ex}}

\newtheorem{lemma}{Lemma}
\newtheorem{definition}{Definition}
\newtheorem{proposition}{Proposition}
\newenvironment{proof}{\medbreak \noindent {\bf Proof.} }{\hfill \rule{2mm}{2mm}
\medbreak}
\newenvironment{remark}{\medbreak \noindent {\bf Remark.} }{\medbreak}
\newcommand{\ex}{\medbreak \noindent {\bf Example. }}

\begin{document}

\title{The symplectic and twistor geometry of the general isomonodromic deformation problem}
\author{
N. M. J. Woodhouse\\The Mathematical Institute, University of Oxford}

\maketitle

\begin{abstract}
\noindent Hitchin's twistor treatment of Schlesinger's equations is extended 
to the general isomonodromic deformation problem. It is shown that a generic 
linear system of ordinary differential equations with gauge group ${\rm 
SL}(n,\C)$ on a Riemann surface $X$ can be obtained by embedding $X$ in a 
twistor space ${\cal Z}$ on which ${\rm sl}(n,\C)$ acts. When a certain 
obstruction vanishes, the isomonodromic deformations are given by deforming 
$X$ in ${\cal Z}$. This is related to a description of the 
deformations in terms of Hamiltonian flows on a symplectic manifold 
constructed from affine orbits in the dual Lie algebra of a loop group. 
\end{abstract}

\subsection*{Introduction}
The study of isomonodromic deformations of systems of ordinary differential 
equations in the complex plane was a significant topic at the beginning of 
the last century, when the classical work of Painlev\'e, Schlesinger, and 
Fuchs was published. It has come back into view in more recent years through 
connections with quantum field theory (Sato {\em et al} 1978, 1979, 1980, 
Dubrovin 1999), differential geometry (Hitchin 1996), and the theory of integrable systems 
(see, for example, Ablowitz and Clarkson 1991).

In this paper, I shall explore in detail one aspect of the modern theory, 
suggested by Hitchin (1995a).  He considered the twistor space of a 
four-dimensional self-dual Riemannian manifold with ${\rm SU}(2)$ symmetry.  
This is a three-dimensional complex manifold ${\cal Z}$, in which there is a 
four-dimensional family of projective lines corresponding to the points of the 
original manifold and on which the symmetry group acts holomorphically. The 
action is generated by three holomorphic vector fields which are 
independent in an open subset, but dependent on a special divisor $S$. By 
taking the vector fields as basis vectors, the tangent space to ${\cal Z}$ is 
identified with ${\rm sl}(2,\C)$ at each point of the open subset.  Thus the 
action determines a flat holomorphic connection on the trivial ${\rm 
SL}(2,\C)$ bundle over the open orbit. 

The restriction of the connection to a twistor 
line is an ${\rm sl}(2,\C)$-valued 
meromorphic 1-form, with poles at the intersections with $S$.  In the case 
that Hitchin considered, there are four poles and the 1-form determines a 
Fuchsian system, with four regular singularities.  As the line is moved 
within the family, the poles move, but the monodromy of the system, which is 
the same as the holonomy of the flat connection, remains unchanged.  By 
calling on the classical theory, therefore, one obtains from this geometrical 
picture a solution to the sixth Painlev\'e equation. Hitchin then goes on to 
exploit this correspondence to construct self-dual Einstein metrics from 
certain Painlev\'e transcendents.

Hitchin's correspondence between twistor manifolds with symmetry and 
iso\-monodromic families of ordinary differential equations holds more 
generally. In this paper, I shall follow through the details of his suggestion 
for the class of isomonodromic deformations considered by Jimbo, Miwa, and 
Ueno (1981). This enables one to understand their results within the framework 
of the general deformation theory of Kodaira (1962). 

In the general setting, 
we are given a complex Lie group $G$, a Riemann surface $X$, 
and a meromorphic 1-form $\alpha$ on $X$ with 
values in the Lie algebra ${\mathfrak g}$. We pick a local coordinate 
$z$ and write $\alpha= -A\, \rd z$.  Then the equation $\rd y+\alpha y=0$ 
becomes a system of linear ordinary differential equations
\be
\frac{\rd y}{\rd z}=Ay,
\label{ODEA}\ee
where $y$ is a fundamental solution, 
taking values in $G$, and $A$ is a meromorphic function of $z$. 
The first question concerns the existence of twistor spaces: this is answered 
by Proposition 1, which gives the existence of an embedding of $X$ in a 
complex manifold ${\cal Z}$ on which ${\mathfrak g}$ acts, with the generators 
independent on an open set, and from which $\alpha$ can be recovered by Hitchin's 
construction. This structure is not unique, however, even if we restrict 
attention to a small neighbourhood of $X$ in ${\cal Z}$.  If $A$ has 
irregular singularities, then there are different choices for the way in which 
a divisor $S$ can be attached to the open set so that the whole of $X$ is 
embedded, including the singularities.  Different choices give different 
possibilities for the normal bundle $N$ of $X$.  
By Kodaira's theorem, the normal bundle determines the deformations of $X$ in 
${\cal Z}$: if $$
H^1(X,N)=0 \quad \mbox{and} \quad \di H^0(X,N)=d_0\, ,
$$ 
then $X$ is one of 
$d_0$-parameter family of compact curves $X_t$, on each of which the ${\mathfrak 
g}$-action gives a linear system of differential equations. These are 
isomonodromic (Proposition 6).  It is shown that there is a natural choice for 
the twistor space (the `full twistor space' in Definition 2), for which the 
parameter space has the largest possible dimension in the generic case; a full 
twistor space exists generically (Proposition 2), and is unique in a 
neighbourhood of $X$ (Proposition 4).  In the full case, $N$ can be 
constructed directly from $\alpha$; if $H^1(X,N)=0$, as is the case if 
$X=\CP_1$, and generally if $\alpha$ has enough singularities, then every 
isomonodromic deformation arises from this construction (Proposition 8).

A second theme of this paper is the Hamiltonian nature of the isomonodromic 
deformation equations.  A {\em Fuchsian system} on $\CP_1$ is a system with 
regular singularities of the form
$$
\frac{\rd y}{\rd z}=\sum \frac{A_iy}{z-a_i}\, ,
$$
where the residues $A_i\in {\mathfrak g}={\rm sl}(2,\C)$ 
are independent of $z$. Apart from gauge 
and coordinate transformations, the only possible deformations in the generic case 
are given by moving the poles $a_i$.  The monodromy is 
then preserved if and only 
if
$$
\frac{\p A_i}{\p a_j}=\frac{[A_i,A_j]}{a_i-a_j}, \quad i\neq j, \qquad
\quad \frac{\p A_i}{\p a_i}=-\sum_{j\neq i}\frac{[A_i,A_j]}{a_i-a_j}\, ,
$$
(Schlesinger's equations).\footnote{One pole is fixed at infinity, and has 
residue $-\sum A_i$.} Hitchin (1997) interpreted these as a Hamiltonian flow 
on the coadjoint orbits of the $A_i$s.  This also generalises:
when irregular singularities are present, the flows are on 
symplectic manifolds constructed from affine orbits in the loop algebra.  The 
symplectic forms can be written down explicitly in terms of $\alpha$ and the
Stokes' matrices, and, at least in the case $X=\CP_1$, one can also find 
explicit expressions for the Hamiltonians (Proposition 9). The symplectic 
structure is related to the structure of ${\cal Z}$ in a neighbourhood of $S$.

An appendix outlines the theory of isomonodromic deformations for linear 
systems on a general Riemann surface.
\bigbreak

\noindent {\bf Acknowledgements.}  This paper draws heavily on ideas on Nigel 
Hitchin as well as on Philip Boalch's D.Phil.\ thesis (on which Boalch 2000 
is based). I thank Majiec Dunajski and Guido Sanguinetti for many interesting 
discussions, and particularly Philip Boalch for his comments and for the corrections he
made to the first draft.

\subsection*{Twistor spaces}

We suppose that we are given a complex Lie group $G$, a Riemann surface $X$, 
and a meromorphic 1-form $\alpha$ on $X$ with 
values in the Lie algebra ${\mathfrak g}$. 
Our starting point is to interpret
a solution $y:X\setminus\{\mbox{poles}\}\to G$ to 
the equation 
\be
\rd y+\alpha y=0
\label{ODE}\ee
as a complex curve in $G$ and to think of $\alpha $ as the pull-back of the 
Maurer-Cartan form on $G$---the ${\mathfrak g}$-valued 1-form whose 
contraction with a left-invariant vector field is the corresponding element of 
${\mathfrak g}$. 

In a local coordinate $z$ on $X$, (\ref{ODE}) is a linear system of ordinary 
differential equations of the form
$$
\frac{\rd y}{\rd z}=Ay\, ,
$$
where $A$ is meromorphic, with values in ${\mathfrak g}$.  Its poles are the 
{\em singularities} of the system---a pole of order $r+1$ is a singularity of 
{\em rank} $r$. Some familiar results about such systems are summarised in the 
appendix.

Of course $y$ is singular at the poles of $\alpha $, and it is multi-valued, 
so the embedding in $G$ is 
defined only locally. In the twistor picture, we seek 
to replace $G$ by a complex manifold ${\cal Z}$ of the same dimension on which 
the right action of $G$ is retained and in which the singular points are 
included. In principle, the construction involves (i) taking a quotient by the 
monodromy group to make $y$ single-valued and (ii) attaching hypersurfaces on 
which the right-action of $G$ action is not free.  The intersections of $X$ 
with the hypersurfaces will then correspond to the poles of the differential 
equation. Except for very special equations, however, the quotient 
is not Hausdorff. The best that we can do in general is to construct ${\cal 
Z}$ as a neighbourhood  of $X$, with the action of $G$ replaced by an action 
of its Lie algebra (which is enough to determine $\alpha $). 
The second step is generally straightforward in the regular case, but is more 
subtle when the linear system has irregular singularities. 

In the context of the deformation problem, we shall adopt a special 
understanding of the meaning of the term `twistor space'. 
\begin{definition} A {\em twistor space} is a complex manifold 
${\cal Z}$ together with
\begin{quote}
(i) a homomorphism from a complex Lie algebra 
${\mathfrak g}$ into the Lie algebra of
holomorphic vector fields on ${\cal Z}$; and

(ii) a smooth compact complex curve $X\subset {\cal Z}$ 
\end{quote}
such that the induced linear map 
$\phi_z: {\mathfrak g}\to T_z{\cal Z}$ is an isomorphism for some 
$z\in X$.
\end{definition}
Note that $\di {\cal Z}=\di {\mathfrak g}$. For the most part, we shall take 
${\mathfrak g}={\rm sl}(n,\C)$, but other examples will also be considered. 

Given a basis in ${\mathfrak g}$, $\Delta=\det \phi$ is a holomorphic section 
of $\bigwedge^{\di {\cal Z}}T{\cal Z}$. We shall make the regularity assumptions  
that 
$$
S=\{\Delta =0\}
$$
is a complex hypersurface, that $X$ is transversal to $S$, that
$S$ is the union of a finite set of components  $S_{i}$, and that  $\Delta$ 
has a zero of order $r_{i }+1$ on $S_{i}$. These hold in all the twistor 
spaces constructed below.

At each $z\in {\cal Z}\setminus S$, define $\theta_z\in T^{*}_z{\cal Z}$ by 
$\theta_z=\phi_z^{-1}$.  Then $\theta$ is a holomorphic 1-form on 
${\cal Z}\setminus S$ with values 
in $\mathfrak g$. It is meromorphic on ${\cal Z}$ and satisfies the {\em 
Maurer-Cartan equation}
\be
\rd \theta+\theta\vp\theta =0
\label{flatness}\ee
Equivalently, $\rd+\theta$ is a flat meromorphic connection on a trivial bundle 
over ${\cal Z}$. The restriction $\alpha =\theta\vert_{X}$ determines a system 
of the form (\ref{ODE}), with poles at the intersection points with $S$. We 
then say that $({\cal Z},X)$ is a {\em twistor space} for the system. 

\ex Let $G={\rm SL}(n,\C)$ and let ${\mathfrak t}\subset {\mathfrak g}={\rm 
sl}(n,\C)$ denote the diagonal subalgebra.  As an $(n-1)$-dimensional additive 
group, ${\mathfrak t}$ acts on itself by translation, and the action extends to 
the compactification $\CP_{n-1}$ when we add a hyperplane at infinity.  
We also have the left action of 
${\mathfrak t}$ on $G$, defined by
$$
g\mapsto \exp(A)g, \qquad A\in {\mathfrak t}.
$$
We put
$$
{\cal Z}=G\times \CP_{n-1}/{\mathfrak t}\, .
$$
Then the right action of $G$ on the first factor descends to the quotient.

We can think of ${\cal Z}$ as being formed by attaching a single hypersurface 
$S$ to $G$ (the projection of the hyperplane at infinity in $\CP_{n-1}$).  The 
effect is to compactify the one-parameter subgroups generated by the 
semisimple elements of ${\mathfrak g}$.  If $A\in {\mathfrak t}$ generates a closed 
subgroup, then $\{tA\}\subset {\mathfrak t}$ compactifies to a projective line in 
$\CP_{n-1}$, and this in turn projects onto an embedded copy of $\CP_1$ in 
${\cal Z}$.  The corresponding system of linear equations is
$$
\frac{\rd y}{\rd z}=Ay\, ,
$$
which has a singularity of rank $1$, a double pole, at infinity (the 
intersection with $S$).

\ex Suppose that $X$ has genus $g$.  Let $G={\cal Z}$ be the Jacobi variety 
(an abelian group with Lie algebra ${\mathfrak g}=\C^g$) and let 
$X\hookrightarrow {\cal Z}$ be the standard embedding (see, for example, 
Farkas and Kra 1980, p.\ 87).  The corresponding system is
$$
\frac{\rd y}{\rd z}=A
$$
where $A=(\xi_1, \ldots ,\xi_g)$, with the $\xi$s a basis for the 
abelian differentials on $X$.  Here there are no singularities. 

\subsubsection*{Existence of twistor spaces}

Does every system of ODEs of the form (\ref{ODE}) have a twistor space?  Since 
the restriction of $\theta$ to $X$ cannot vanish, a necessary condition is 
that
$\alpha \neq 0$ at every point of $X$. This condition holds in the generic 
case (since it fails only if the all the entries in the matrix $A$ have a 
coincident zero). It is also sufficient.
\begin{proposition}  
Let $\alpha $ be a meromorphic ${\mathfrak g}$-valued 1-form on $X$ with no 
zeros. Then the linear system of ODEs $\rd y+\alpha y=0$ has a twistor space. 
\label{minprop}
\end{proposition}
\begin{proof} 
We construct ${\cal Z}$ by taking a quotient of a neighbourhood of the 
identity section $X$ in $X\times G$ by a distribution $F$ constructed from 
the linear system. 

Let $D$ be a neighbourhood of a pole $a$ not containing any of the 
other poles, and let $z$ be a coordinate on $D$ such that $z=0$ at $a$.
Then $\alpha=-A\,\rd z$ in $D$, where $A:D\setminus 
\{0\}\to {\mathfrak g}$ is holomorphic and has a pole of order $r+1$ at $z=0$.

Define $F$ to be the distribution on $D\times G$ tangent to the 
non-vanishing vector field 
$$
z^{r+1}(\p_{z}-R_A)
$$
where $R_A$ is the right-invariant vector field on $G$ generated  by $A(z)$.  If 
$D'$ is an open set not containing any other poles, then we define $F$ in the 
same way on $D'\times G$, but without the factor $z^{r+1}$; that is $F$ is 
tangent to $\p_z-R_A$. The vector fields are proportional on $D\cap D'$, so 
$F$ is well defined globally as a distribution on $X\times G$. Under the 
condition on $\alpha$, we have $F\cap T_xX=0$ at every $x\in X$. So it is 
possible to choose an open neighbourhood $N$ of $X$ in $X\times G$
such that the quotient 
${\cal Z}=N/F$ is a Hausdorff complex manifold of the same dimension as $G$.  
We then have a double fibration

\setlength{\unitlength}{1.3mm}
\begin{center}
\begin{picture}(28,15)
\put(2,0){$\CP_1$}
\put(24,0){${\cal Z}$}               
\put(13,10){$N$}
\put(11,8){\vector(-1,-1){5}}
\put(6,7){$\pi_1$}
\put(20,7){$\pi_2$}
\put(17,8){\vector(1,-1){5}}
\end{picture}
\end{center}
\medbreak

\noindent and a smooth curve $\pi_2(X)\subset {\cal Z}$,
which we also denote by $X$.

Because we are looking only at a neighbourhood of the identity section, the 
right action of $G$ on $X\times G$ does not pass to ${\cal Z}$; but the 
corresponding Lie algebra action does.  Each $v\in {\mathfrak g}$ can be 
identified with a left-invariant vector field on $G$, and hence with a vector 
field on $X\times G$ tangent to the fibres of $\pi_1$. Its projection 
$V$ by $\pi_{2{*}}$ is a holomorphic vector field on ${\cal Z}$, and the map 
$v\mapsto V$ is a Lie algebra representation, satisfying the conditions in 
the definition of a twistor space. The singular hypersurface has components 
given by the poles of $\alpha$, and $X$ meets these transversally.
                                                 
It remains to show that  $\theta\vert_{X}=\alpha $. To do this, we note that 
the meromorphic 
vector field $\p_z-R_A$ on $\CP_1\times G$ is tangent to $F$, and so its 
projection into ${\cal Z}$ vanishes.  On the other hand, at the identity, 
the 
right- and left-invariant vector fields generated by an element of ${\mathfrak 
g}$ coincide. Hence $i^{\str}_{\p_z} \theta =-A$. The proposition follows. 
\end{proof}
\begin{remark}
If we instead take the curve in ${\cal Z}$ to be the projection under $\pi_2$ 
of $X\times \{g\}$ for some other constant element of $g$, then we obtain 
instead a twistor space for $g^{-1}\alpha  g$.
\end{remark}

In the irregular case, the twistor space is not unique: the one that 
arises in Proposition \ref{minprop} is {\em minimal} in a sense that will 
be explained later.

\subsubsection*{Full twistor spaces}

The difference in structure between different twistor spaces of a system of 
ODEs can be understood by looking at the structure in a neighbourhood of a 
point of $a\in S$. By introducing a local coordinate $z$ that vanishes on 
$S$, we can choose a neighbourhood $U$ of the form $S\times D$, where 
$D\subset \C$ is, say, the unit disc, and the $X\cap U$ is $\{(a,z)\}$, 
$z\in D$. 

Suppose that $\alpha=-A\, \rd z$ has a pole of order $r+1$ at 
$z=0$.
Then corresponding system 
\be
\frac{\rd y}{\rd z}=Ay
\label{Azero}\ee
has a singularity of rank $r$ at $z=0$. 

Given a holomorphic vector field $V$ on ${\cal Z}$ tangent to $S$, we can 
construct a holomorphic family of copies $D_t$ of $D$ in $U$ 
by moving $D_0=X\cap U$ along $V$ (and if 
necessary restricting to a smaller neighbourhood of $a$):  here 
$t$ is a complex 
parameter taking values in some neighbourhood of $t=0$.
By 
restricting $\theta$ to each $X_t$, we get a one-parameter family of ODEs 
$$
\frac{\rd y}{\rd z}=A(z,t)y\, ,
$$
each with a singularity of rank $r$ at $z=0$ (the singularity does not move 
with $t$ because $V$ is tangent to $S$).  It follows from
the flatness of $\rd 
+\theta$ that
$$
\frac{\p A}{\p t}=\frac{\p \Omega}{\p z}-[A,\Omega]\, ,
$$
where $\Omega=-i^{\str}_V \theta$. This is the {\em local deformation equation} 
(see appendix).  

At each fixed value of $t$, $\Omega$ is a function of the coordinate $z$.
Introducing the notation
$$
\rD\Omega=\p_z\Omega- [A,\Omega]
$$
we have
\be
\Omega= O(z^{-r-1}), \qquad \rD\Omega=O(z^{-r-1})\quad \mbox{as $z\to 0$.}
\label{Omcond}\ee
When the singularity is irregular, the various twistor spaces differ in the 
extent to which the converse holds: in the minimal construction, an $\Omega$ 
satisfying these conditions is of the form $i^{\str}_V \theta$ for some 
holomorphic $V$ only if $\Omega - f(z) A$ is holomorphic at $z=0$ for some 
holomorphic function $f$.  
\medbreak
\begin{definition} 
A twistor space ${\cal Z}$ is {\em full} at $a\in S$ 
if for every $\Omega:D\to {\mathfrak g}$ such that (\ref{Omcond}) holds, 
there is a holomorphic vector field $V$ on $U\subset {\cal Z}$ such that 
$\Omega=-i^{\str}_V \theta\vert_X$.  
The twistor space is {\em full} if it is full at 
every point of $S$.
\end{definition} 
When the system has irregular singularities, and Rank$(G)>2$, the twistor 
space constructed in Proposition (\ref{minprop}) is not full. We can see this 
by noting that, for any holomorphic $V$, $i_V\theta$ has singular part at each 
pole that is proportional (by a holomorphic function) to a multiple of $A$, 
and cannot therefore give rise to the most general $\Omega$ satisfying 
(\ref{Omcond}). We shall put this more precisely below when we consider the 
normal bundle of $X$ in ${\cal Z}$.

A full twistor space generates not only the ODE itself, but also its 
isomonodromic deformations. 
We shall see that it is possible to construct a full twistor space in the 
generic case, but there are some rather special exceptions.
A necessary condition for existence is that if 
$\Omega$ and $\Omega'$ 
both satisfy (\ref{Omcond}), and if 
$$
\Omega=\frac{M}{z^{r+1}}+O(z^{-r}), \qquad
\Omega'=\frac{M'}{z^{r+1}}+O(z^{-r})\, ,
$$
as $z\to 0$, 
with $k>0$, then $[M,M']=0$.  This fails in the following class of examples. 

\ex Suppose that
$$
A=z^{-2}\pmatrix{1&0\cr 0&1\cr}\, ,
$$
Then 
$$
\Omega=\frac{M}{z}, \qquad
\Omega'=\frac{M'}{z}\, ,
$$
satisfy (\ref{Omcond}) for any constant $M$, $M'$; but 
both cannot be generated 
from vector fields in the same twistor space if $[M,M']\neq 0$.
\medbreak

\subsubsection*{The singular hypersurface}
                              
We now look in detail at the structure of a twistor space in a neighbourhood 
of $S$. The fullness condition at a point of $X\cap S$ is this: given $\Omega$ 
such that (\ref{Omcond}) holds, is $\phi(\Omega)$ holomorphic at $z=0$? 

Let us write
\be
A=\frac{p}{z^{r+1}}+H,\qquad \Omega=\frac{\omega}{z^{r+1}}
\label{normform}\ee
where 
$$
p=p_0+p_1\, z+\cdots + p_r\, z^r, \qquad
\omega=\sum_0^{\infty}\omega_iz^i
$$
and $H$ is holomorphic on $U$. If $\omega=O(z^{r+1})$ as $z\to 0$, then 
$\Omega$ is holomorphic at $z=0$ and can be generated by the  holomorphic 
vector field $\phi(\Omega)$ in any twistor space for $A$; so the source of any 
difficulty lies in the first $r$ terms in the Taylor expansion of $\omega$.

When we separate out the 
coefficients of $z^{-2(r+1)}$, $z^{-2r-1}$, \ldots $z^{-(r+2)}$ 
in (\ref{Omcond}), we obtain for $r>0$
\begin{eqnarray*}
[p_0,\omega_0]&=&0\\
{}[p_0,\omega_1]+[p_1,\omega_0]&=&0\\
&\vdots&\\
{}[p_0,\omega_{r-1}]+[p_1,\omega_{r-2}]+\cdots +[p_{r-1},\omega_0]&=&0\\
{}[p_0,\omega_{r}]+[p_1,\omega_{r-1}]+\cdots 
+[p_{r},\omega_0]-(r+1)\omega_0&=&0\, 
,\\
\end{eqnarray*}
or in the case $r=0$,
\be
[p_0,\omega_0]=\omega_0\, .
\label{adjeig}\ee
The generic case is that
one or other of the following hold:
\begin{quote} 
(i) $r>0$ and the eigenvalues of $p_0$ are distinct; or

(ii) $r=0$, the eigenvalues of $p_0$ are distinct, and no pair differ by an 
integer.\footnote{To prove Lemma \ref{lemm1} and Proposition \ref{fullprop}, 
we only need that no pair should differ by 1;  however the stronger condition 
here is needed to construct $g^{\str}_{{\rm f}}$ (see Appendix), and is 
imposed here to avoid special cases in the presentation below.}
\end{quote}
\begin{lemma}
If $A$ is generic, then $\Omega$ satisfies (\ref{Omcond}) if and only if
$\omega_0=0$ and $[\omega,p]=O(z^{r+1})$ as $z\to 0$.
\label{lemm1}\end{lemma}
\begin{proof}
Under either of the conditions (i), (ii), the eigenvalues of $p$ can be 
assumed to be distinct, since they are distinct at $z=0$ and since we can, if 
necessary, replace $U$ by a smaller neighbourhood. So we can find a 
holomorphic gauge transformation $g:U\to G$ such that $g^{-1}pg$ is the sum of 
a diagonal polynomial and a term that vanishes to order $z^{r+1}$ at $z=0$, 
and so can be absorbed into $r$. 

If we assume first that $p$ is actually diagonal, then 
we deduce successively that $\omega_0,\ldots \omega_{r-1}$ are diagonal (for 
$r>0$) and that 
$$
[p_0,\omega_r]=(r+1)\omega_0\, .
$$
For $r>0$, this implies that $\omega_0=0$ since the diagonal terms on the 
left-hand side vanish, and hence that $\omega_r$ is also diagonal.  
For $r=0$, it gives $\omega_0=0$ since $p_0$ has no pair of eigenvalues 
differing by 1. Thus, whether or not  $p$ is diagonal, we have
that $\Omega$ is holomorphic at $z=0$ when $r=0$; and that when $r>0$,
$$
\Omega= \frac{g q g^{-1}}{z^r}+O(z^{0})
$$
where $q$ is a diagonal polynomial of degree $r-1$. 
\end{proof}

\begin{proposition} Let $\rd y+\alpha  y=0$ be a generic system, with $G={\rm 
SL}(n,\C)$.
Then there exists a full twistor space.\label{fullprop}
\end{proposition}
\begin{proof}
Any twistor space is full at a singularity of rank $r=0$ since any $\Omega$ 
satisfying (\ref{Omcond}) is then holomorphic at $z=0$, and can therefore be 
generated by a holomorphic vector field in any twistor space. In the irregular 
case, we construct ${\cal Z}$ from the `minimal' twistor space in Proposition 
\ref{minprop}, by cutting out and replacing a neighbourhood of each component 
of $S$ corresponding to an irregular singularity. 

Suppose, to begin with, that the system has a singularity of rank $r$ at $z=0$ 
and that in a neighbourhood $D$ of $z=0$ we have $\alpha=-A\, \rd z$, where
\be
A=\frac{p}{z^{r+1}}+H
\label{diagA}
\ee
with $p$ a diagonal polynomial of degree $r$ with distinct diagonal 
entries throughout $D$ and $H$ holomorphic. By making a diagonal gauge 
transformation, we can make $H$ off-diagonal.  

Pick constant diagonal matrices $q_1, \ldots ,q_{n-2}$ which, 
together with $p(0)$, form a basis for the diagonal subalgebra of ${\mathfrak g}$, 
and for each $i$ let $H_i$ be the off-diagonal matrix with entries 
$$
(H_i)_{ab}=\frac{zH_{ab}(q_{ia}-q_{ib})}{p_a-p_b}, \qquad a\neq b\, ,
$$
where $q_{ia}$ and $p_a$, $a=1, \ldots, n$, are the diagonal entries in $q_i$ 
and $p$. Thus \be [p,H_i]=[q_i,zH], \qquad
[q_i,H_j]=[q_j,H_i]\label{pqH}\, .\ee
Now introduce evolution equations for 
the diagonal matrix 
$p(z)$ and the off-diagonal matrix $H(z)$ as functions of the complex 
variables $t_1, \ldots, t_{n-1}$ by putting
$$
\p_i p=-rq_i
\qquad
\p_i H=\p_z H_i-[H,H_i]\, ,
$$
where $\p_i=\p/\p t_i$, $\p_z=\p/\p z$.
The integrability of this system is established by showing that
$$
\p_iH_j-\p_jH_i=[H_i,H_j]\, .
$$
Since both sides are off-diagonal, this follows from (\ref{pqH})  and
\begin{eqnarray*}
[p,\p_iH_j]-[p,\p_jH_i]&=&z[q_j,\p_iH]-z[q_i,\p_jH]\\
&=&z[q_j,\p_zH_i]-z[q_i,\p_zH_j]-z\bigl[q_j,[H,H_i]\bigr]
+z\bigl[q_i,[H,H_j]\bigr]\\
&=&z\bigl[H_i,[q_j,H]\bigr]-z\bigl[H_j,[q_i,H]\bigr]\\
&=&\bigl[p,[H_i,H_j]\bigr]\, .
\end{eqnarray*}
So the evolution equations 
extend $H$ and $p$ to functions of $(z,t_1,\ldots t_{n-2})$ on a neighbourhood 
$W$ of the origin in $\C^{n-2}$. 

It follows from the definitions that
$$
\nabla=\rd - \frac{p\, \rd z}{z^{r+1}} -H\,\rd z-\sum_i\left(\frac{q_i\, \rd 
t_i}{z^r} +H_i\, \rd t_i\right)
$$
is flat meromorphic connection on the trivial bundle principal bundle 
$P=G\times W$.

Let ${\cal Q}$ denote the quotient of a neighbourhood of the identity section 
in $P$ by the horizontal foliation. The foliation 
extends holomorphically to $z=0$ since it is spanned by
$$
z^{r+1}\p_z-p-z^{r+1}H, \quad z^r\p_i-q_i-z^rH_i\quad (i=1, \ldots ,n-2)\, ,
$$
where $p,q_i,H,H_i$ are interpreted as right-invariant vector fields on $G$.

The quotient is a `local twistor space' for $A$ in the sense that it 
carries a holomorphic ${\mathfrak g}$-action, which is free and transitive except 
on the hypersurface $S'=\{z=0\}$, and  contains a copy of $D$ on which the 
induced system is $\alpha$.  
Moreover, the fullness condition holds at $S$ since 
$p_0$ and the $q_i$s span the diagonal subalgebra (in the case $n=2$, $p_0$ 
on its own does that). Any generic $A$ can be 
reduced to the form (\ref{diagA}) by a holomorphic 
gauge transformation $g(z)$; so more generally a local twistor space can be 
constructed by applying the same gauge transformation to $\nabla$. 

By using the ${\mathfrak g}$ action, we can identify ${\cal Q}\setminus S'$ with 
$V\setminus S$, where $V$ is a neighbourhood in ${\cal Z}$ of the $z=0$ 
intersection point of $S$ and $X$. Then the embedded copy of 
$D\setminus \{0\}$ is mapped onto 
a punctured neighbourhood of the singularity in 
$X$.  The identification allows us to replace $V$ by ${\cal Q }$. By 
repeating this for the other irregular singularities, we obtain a full twistor 
space. 
\end{proof}

Given the choice of coordinate $z$ in a neighbourhood of an irregular 
singularity, $p_0$ is a well-defined map $S\to {\mathfrak g}$.  Up to scale, $p_0$ 
is independent of the choice of $z$. We thus have a natural map $[p_0]:S\to 
{\mathbb P}{\mathfrak g}$.  It is equivariant with respect to the ${\mathfrak g}$ 
action on $S$ and the adjoint action on ${\mathbb P}{\mathfrak g}$. The following 
is immediate from the proof above. 

\begin{proposition} A twistor space for a generic system 
is full at an irregular singularity 
if and only if $[p_0]:S\to {\mathbb P}{\mathfrak g}$ 
is regular at $X\cap S$.
\end{proposition} 
When the space is not full, $[p_0](S)$ has nonzero codimension in ${\mathbb 
P}({\mathfrak g})$. In the generic case, the full twistor space is locally 
unique in the sense of the following proposition. 

\begin{proposition} Suppose that $({\cal Z},X)$ and $({\cal Z}',X)$ are full 
twistor spaces for a generic linear system of ODEs on a Riemann surface $X$, 
with $G={\rm SL}(n,\C)$.  Then there are neighbourhoods $U\supset X$ and 
$U'\supset X$ in ${\cal Z}$ and ${\cal Z'}$ and a ${\mathfrak g}$-equivariant 
biholomorphic map $\rho:U\to U'$ such that $\rho(X)=X$. 
\label{globunique}\end{proposition} 
If we exclude the poles from $X$ and the corresponding hypersurfaces $S$ and 
$S'$ from ${\cal Z}$ and ${\cal Z'}$, then $\rho$ is determined in a 
straightforward way by the ${\mathfrak g}$ actions on ${\cal Z}$ and ${\cal Z}'$. 
It is defined by choosing a (multivalued) solution $y$ to the system on $X$ 
and then extending $y$ to a (multivalued) map from a neighbourhood of 
$X\setminus S$ in ${\cal Z}\setminus S$ to $G$ such that $\rd y+\theta y=0$.  
Similarly $y$ extends to $y'$ on ${\cal Z}'$.  Then the required map is 
$\rho=y^{\prime -1}\circ y$, where the branches are chosen so that $\rho$ is 
the identity on $X$ ($\rho$ is well defined since $y$ and $y'$ have the same 
holonomy). The fact that $\rho$ extends holomorphically to $S$ in the full 
case is a corollary of Proposition~\ref{locunique} below. 

\subsubsection*{The structure of S}

Suppose that $G={\rm SL}(n,\C)$ and that $({\cal Z},X)$ is full. 
We denote by $\Gamma$ the space of 
parametrized curves 
$$
D\to {\cal Z}:z\mapsto \gamma(z),
$$
where $D\subset \C$ is the unit disc,
$\gamma(D)$ meets some component of $S$
transversally at $z=0$, and $\gamma$ extends smoothly to $\vert z\vert=1$.  

We shall now develop a picture in which $\theta_{\gamma}=\gamma^*(\theta)$ is 
regarded as  an element of the dual of Lie algebra of the loop group 
$LG$.\footnote{An element of loop algebra is a map $B:S^1\to {\mathfrak g}$. We
define $\langle\theta_\gamma,B\rangle$ by integrating $\tr(B\theta_\gamma)$ 
around the unit circle.} Different elements of $\Gamma$ give different points 
of an orbit in $L{\mathfrak g}^{{*}}$ of an affine action of $LG$. We shall 
construct a finite-dimensional complex symplectic manifold from the orbit 
which determines the singular part of $\alpha$ at $z=0$. 

\subsubsection*{The Fuchsian case}

If the singularity at $z=0$ is regular and generic, then any twistor space is 
full at $z=0$ and any holomorphic map $z\to \Omega(z)\in {\mathfrak g}$ generates 
a local holomorphic vector field tangent $Y$ to $S$.  
Moreover we can write
$$
\theta = p_0\, \rd(\log z)+\theta'
$$
where $z=0$ on $S$ and $\theta'$ is holomorphic on $S$.\footnote{Thus $\theta$ 
has a logarithmic pole in the sense of Malgrange (1982).} Then $p_0$
is independent of the choice of the 
function $z$, and therefore determines a natural map $\mu:S\to {\mathfrak 
g}={\mathfrak g}^{*}$---the identification being given by the bilinear form  
$\tr(\xi_1,\xi_2)$, $\xi_i\in {\mathfrak g}$. The image $\mu(S)$ is open subset of 
a coadjoint orbit. 

By evaluating $\Omega$ and $Y$ at $z=0$, we obtain a natural identification 
$$
T_aS={\mathfrak g}/[p_0(a)]\qquad a\in S\, .
$$
We can therefore define a 2-form $\sigma$ on 
$S$ by
$$
\sigma_a(Y,Y')=\tr(p_0[\Omega,\Omega'])\, .
$$
This is closed and presymplectic since it is the pull back to $S$ by $\mu$ of 
the symplectic form on $\mu(S)$. 

\subsubsection*{The irregular case}

In the irregular case, the analogous structure involves information from the 
higher formal neighbourhoods of $S$. It arises from the action 
on $\Gamma$ of the group 
$L_+G$ of holomorphic maps $g:D\to G$ that extend smoothly to $\ov 
D$: if $g\in L_+G$ then 
$\gamma\in \Gamma$, then $(g\gamma)(z)=\gamma(z)g(z)$.\footnote{Of course this 
is well-defined only for $g$ close to the identity; what follows has to be 
qualified in a similar way.}

A tangent vector $Y$ to $\Gamma$ at $[\gamma]$ is a 
section of $T{\cal Z}\vert_{\gamma}$, tangent to $S$ at $z=0$. 
Put
\be
\sigma_{\gamma}(Y,Y')=\frac{1}{2\pi\ri}\oint \tr(\Omega \nabla\Omega')\, ,
\label{sigdef}\ee
where $\Omega=i_{Y}\theta$, $\Omega'=i_{Y'}\theta$,
$\nabla\Omega=\rd \Omega +[\theta,\Omega]$ 
and the integral is along a loop surrounding 
$z=0$. This form is closed, but degenerate (its closure follows from the 
construction below). Its characteristic distribution is integrable, by closure, and 
contains the $Y$s 
for which $\Omega=O(z^{r+1})$ as $z\to 0$. These span 
the orbits of the normal subgroup $L_{r+1}G\subset L_+G$ of maps 
$g:D\to G$ such that $g=1+O(z^{r+1})$ as $z\to 0$. Since $\Gamma/L_{r+1}G$ is finite-dimensional, 
the quotient $\Gamma_r$ of $\Gamma$ by the characteristic distribition is a finite-dimensional 
symplectic manifold.

Since ${\cal Z}$ is full, the tangent space to $\Gamma_r$ at  $[\gamma]$ is 
the set of holomorphic maps 
$$
\Omega:D\setminus \{0\} \to {\mathfrak g}
$$
such that (\ref{Omcond}) holds, modulo maps with zeros of order $r$ at $z=0$.

\subsubsection*{Affine orbits}

Let $LG$ denote 
the loop group of smooth maps $S^1\to G$ (Pressley and Segal, 1986). 
Its Lie algebra is the space $L{\mathfrak g}$ of smooth maps 
$\Omega:S^1\to {\mathfrak g}$. 

A $1$-form $\alpha$ on $S^1$ with values in ${\mathfrak g}$ determines an element 
of $L{\mathfrak g}^{{*}}$ by
$$
\langle \alpha, \Omega\rangle=\frac{1}{2\pi\ri}
\oint \tr (\Omega\,\alpha)
$$
Let ${\cal A}_r\subset L{\mathfrak g}^{{*}}$ denote the subspace of smooth
$\alpha$s that 
extend meromorphically to $D$ with a pole at the origin of order at most 
$r+1$, and no other poles. Thus 
$$
L{\mathfrak g}^{{*}}\supset {\cal A}_r\supset {\cal A}_{r-1}\supset \cdots \supset
{\cal A}_0\, .
$$

Like any dual Lie algebra, $(L{\mathfrak g})^{*}$ carries the standard {\em 
Kostant-Kirillov-Souriau} Poisson structure, which is preserved by the 
coadjoint action of $LG$. The natural symplectic arena for the isomonodromy 
problem is not, however, that of the corresponding coadjoint orbits, but 
rather that of the orbits of the {\em affine action} of $L{\mathfrak g}$ on 
$L{\mathfrak g}^{*}$ given by the cocycle 
$$
c(\Omega,\Omega')=\frac{1}{2\pi \ri}\oint \tr(\Omega\, \rd \Omega')\, .
$$
Each $\Omega\in L{\mathfrak g}$ determines a vector field on $L{\mathfrak g}^{*}$, its 
value at $A\in L{\mathfrak g}^{*}$ being
given by
$$
\langle\delta \alpha , \cdot \rangle = \langle \alpha, [\Omega, \cdot]\rangle 
-c(\Omega,\,\cdot\,)\, .
$$
The first term on the right-hand side is the usual coadjoint action; the second 
is a translation introduced by Souriau 
(1970) (see also Woodhouse 1990).\footnote{For a general Lie group, 
the affine action is symplectic, but not Hamiltonian 
when $c$ is not a coboundary. In fact the affine orbits are the models for the 
non-Hamiltonian transitive symplectic actions of the group in the same way 
that the coadjoint orbits are the models for the Hamiltonian actions. If $c$ 
is the obstruction to the existence of a moment map for some transitive 
symplectic action of the Lie algebra on a symplectic manifold $M$, then $M$ can 
be mapped equivariantly and symplectically to an affine orbit in the dual Lie 
algebra.  The affine orbits have an alternative description in terms of the 
coadjoint orbits of the central extension determined by $c$, but this is less 
convenient for our purposes here. See Pressley and 
Segal (1986), p.\ 44.} 
By integrating the flow on 
$L{\mathfrak g}^{{*}}$, we obtain the gauge action of $LG$:
$$
\alpha \mapsto g^{-1}\alpha g+g^{-1}\rd g\, .
$$
The symplectic structure on the corresponding orbits is\footnote{
This is a good definition of $\sigma$ since the right-hand side vanishes 
whenever $\Omega_1$ fixes $A$, and therefore whenever $Y_1$ vanishes at $A$. 
We shall not need to consider the precise sense in which $\sigma$ is 
nondegenerate, since we shall deal only with finite-dimensional submanifolds.} 
\be
\sigma(Y,Y')=c(\Omega,\Omega')-\langle \alpha, [\Omega,\Omega']\rangle 
=\frac{1}{2\pi\ri}\oint\tr\bigl(\Omega\nabla\Omega'\bigr)
=\frac{1}{2\pi\ri}\oint\tr\bigl(\Omega\delta' \alpha\bigr)\, ,
\label{sigY}\ee
where $Y,Y'$ are the vector fields generating the actions of 
$\Omega,\Omega'\in L{\mathfrak g}$, $$\nabla\Omega=\rd 
\Omega+[\alpha,\Omega]\, ,$$ 
and $\delta' \alpha$ is the variation induced by $Y'$.
The flow of $\Omega \in L{\mathfrak g}$ is generated 
by 
$$
h(\alpha)=\frac{1}{2\pi\ri}\oint \tr (\alpha\Omega)\, .
$$
However, $[h_A,h_B]=h_{[A,B]}+c(A,B)$, so the action is not Hamiltonian: the 
cocycle is the obstruction to the existence of a moment map.

\subsubsection*{The symplectic structure of $M_r$}
Let ${\cal O}$ be the affine orbit of some generic element of 
${\cal A}_r$; that is, an element of the form
$$
g^{-1}\,\rd g- g^{-1}\left(\frac{p}{z^{r+1}}+H\right)g\, ,
$$
where $g\in L_+G$, $H\in L_+{\mathfrak g}$ are holomorphic on $D$, 
and $p$ is a polynomial in $z$ with distinct eigenvalues for $z\in D$.

A general element $\alpha \in {\cal O}$ is a smooth 1-form on $S^1$ with 
values in ${\mathfrak g}$.  For any $\alpha,\alpha'\in {\cal O}$, the 
two systems
$$
\rd y+\alpha y=0, \qquad 
\rd y+\alpha' y=0
$$
on the circle have the same monodromy matrix $M$ up to conjugacy, since that 
is the condition that $y$ and $\hat y$ can be chosen so that $g=y(z)\hat 
y(z)^{-1}$ is single valued; $g$ is then the element of $LG$ that maps one 
system into the other. 

Since $c$ vanishes on $L_+{\mathfrak g}$, the action of $LG$ on ${\cal O}$ 
{\em is} Hamiltonian when 
restricted to $L_{r+1}G$, the subgroup of loops of the form $1+z^{r+1}h$, 
where $h$ is holomorphic. We denote by ${\cal M}_r$ the Marsden-Weinstein 
reduction of $\mu_{r+1}^{-1}(0)$, where $\mu_{r+1}$ is the moment map.  That 
is, ${\cal M}_r$ is the quotient of ${\cal A}_r\cap {\cal O}(A)$ by the action 
of $L_{r+1}G$.  It a finite-dimensional complex symplectic manifold, with 
symplectic form that we shall again denote by $\sigma$. Its points are 
elements of ${\cal A}_r\cap {\cal O}(A)$, modulo gauge transformations of the 
form 
$$
\alpha \mapsto g^{-1}\, \rd g- g^{-1}\alpha g, \qquad g\in L_{r+1}G\, .
$$                  
A tangent vector $Y\in T_{\alpha}{\cal M}_r$ is of the form
$$
\delta \alpha =\nabla \Omega= \rd \Omega +[\alpha,\Omega]
$$
where $\Omega$ satisfies (\ref{Omcond}); two such $\Omega$s define the same 
tangent vector whenever their difference is in  $L_{r+1}{\mathfrak g}$.

We can construct from $\alpha\in {\cal A}_r$
the following objects (see appendix).
\begin{quote}
(i) The singularity data $(m,t)$, where $t$ is a diagonal polynomial of degree
$r-1$ and $m$ is the exponent of formal monodromy.

(ii) The formal series $g^{\str}_{{\rm f}}=\sum g_iz^i$.

(iii) The matrices $C_{i}$, defined as follows. For each $\alpha$, choose a 
solution $y$ to the corresponding system with fixed monodromy matrix $M$ and 
choose $2r$ sectors ${\cal S}_{i}$ at $z=0$, as in (\ref{sect}). Then put 
$C_{i}=y_{i}(z)^{-1}y(z)$, where $y_{i}$ is the corresponding special solution 
and we continue $y$ in the positive sense around $z=0$ into the sector ${\cal 
S}_i$. If we put
$$
y_{2r+1}=y_1\re^{2\pi\ri m}\qquad {\cal S}_{2r+1}={\cal S}_{1}\, , 
$$
then $C_{2r+1}=\re^{-2\pi\ri m}C_1M$ and we can define the {\em Stokes' 
matrices} by $S^{\str}_{i}=C^{\str}_{i}C_{i+1}^{-1}$ ($i=1,\ldots 2r$). 
\end{quote} 
These are not quite uniquely determined: we are free 
to make the replacement
\be
g^{\str}_{{\rm f}}\mapsto
g^{\str}_{{\rm f}}T, \qquad 
C_{i}\mapsto T^{-1}C_{i}, \quad S_i \mapsto T^{-1}S_iT\, ,
\label{freedom}\ee
where T is diagonal and independent of $z$.
We shall express the symplectic form on ${\cal M}_r$ in terms of these 
variables.

Given $t$ and the monodromy matrix $M$, the Stokes' matrices and the exponent 
of formal monodromy satisfy two constraints. 
\begin{quote}
(C1) Let $P_i$ be the matrix of the permutation that puts the real parts of 
diagonal entries in $z^{-r}t$ is increasing order
in ${\cal S}_i\cap {\cal S}_{i+1}$.  Then for each $i$,
$P_i^{-1}S_iP_i$ is upper triangular and $P_iS_{i+1}P_i^{-1}$ is lower triangular, 
both with ones on the diagonal.  This follows 
from the fact that
\be
\exp (z^{-r}t+m\log z)S_{i}
\exp (-z^{-r}t-m\log z)\to 1
\label{Stokeasym}\ee
faster than any power of $z$ 
as $z\to 0$ in ${\cal S}_i\cap {\cal S}_{i+1}$. 

(C2) The product $\re^{-2\pi\ri m}S_1\ldots S_{2r}$ is conjugate to $M^{-1}$.
\end{quote}
Denote by ${\cal C}_r$ the set matrices $S_i\in {\rm SL}(n,\C)$, $m$ diagonal and trace-free, 
satisfying these two constraints.  
Given a point ${\cal C}_r$, we choose $C_1$ such that
$$
\re^{-2\pi\ri m}S_1\ldots S_{2r}=C_1^{\str}M^{-1}C_1^{-1}
$$
and define $C_2, \ldots, C_{2r+1}$ by $C_{i+1}=S_i^{-1}C_i$.  Put
\be
\omega=\frac{1}{2\pi \ri}\sum_1^{2r}\tr \Bigl(\rd
C^{\str}_iC_i^{-1}\otimes \rd S^{\str}_i S^{-1}_i\Bigr) +\pi\ri\, \tr\bigl(\rd m 
\otimes\rd m) -\tr(\rd C^{\str}_1C_1^{-1}\otimes\rd m)\, .
\label{omdef}\ee
It is shown in the appendix that $\omega$ is skew-symmetric, and in fact a symplectic form
on ${\cal C}_r$.

For each point of ${\cal M}_r$, we pick a representative in $\alpha\in
{\cal A}_r$.  We then define 1-forms 
on ${\cal M}_r$ by
$$
\Theta= \rd g^{\str}g^{-1} +
\frac{g\rd t\,g^{-1}}{z^{r}}
\qquad \gamma = g_0^{-1}\rd g_0
\, ,
$$
where $\rd$ is the exterior derivative on ${\cal M}_r$ and
$g$ is the polynomial obtained by truncating the formal power series $g_{{\rm 
f}}$ at some large power of $r$. 
With this notation, the symplectic form on 
${\cal M}_r$ is given by the following proposition.
\begin{proposition} The symplectic form on ${\cal M}_r$ is
$$\sigma=
\frac{1}{2\pi\ri}\oint\tr\bigl(\Theta\vp \nabla\Theta'\bigr)
+\tr\bigl(\gamma \vp \rd m\bigr)-\omega\, .
$$
where 
$\nabla\Omega=\p_z\Omega\, \rd z +[\alpha,\Omega]$.
\end{proposition} 
The proof is by splitting the integral in the definition of $\sigma$ into sections 
lying in the various sectors, and then shrinking the contour to zero.
The details are given in the appendix.

The formula for $\sigma$ is independent of the choices made in defining the variables on ${\cal M}_r$.
In particular, it depends on the first $r$ terms in the formal series since the right-hand 
side of the formula is unchanged when $\Theta$ is replaced by $\rd hh^{-1}+h\Theta h^{-1}$, where 
$h=1+O(z^{r+1})$. It is also unchanged when $g$, $C_i$ are replaced by $gT$, $T^{-1}C_i$ where $T$
is diagonal and indpendent of $z$.

We can see the local structure of ${\cal M}_r$ from the proposition.  The 
submanifolds on which $g$ and $t$ are constant (up to the freedom \ref{freedom}) are symplectomorphic to ${\cal C}_r$.
While, those on which $m$ and $S_i$ are constant (up to \ref{freedom}) are symplectomorphic to a fixed
manifold ${\cal P}_r$; by mapping $[\gamma]\in \Gamma_r$ to $[\alpha]$, where $\alpha=\gamma^{*}(\theta)$, 
and by noting the coincidence of the formulas for the symplectic forms, we identify ${\cal P}_r$ with
$\Gamma_r$. 

\subsection*{Local uniqueness of the full twistor space}

Let ${\cal P}$ denote the subset of ${\cal O}\cap {\cal A}_r$ 
given by the fixing the values of the Stokes' matrices and exponent
of formal monodromy. By using the actions of $L_{r+1}G$ on $\Gamma$ 
and ${\cal P}$ to pick represenetatives in $[\gamma]$ and $[\alpha]$,
we can identify $\Gamma$ with an open neighbourhood in ${\cal P}$ so 
that $\gamma\in \Gamma$ corresponds to $\alpha\in {\cal P}$ such at 
$\alpha=\gamma^*(\theta)$. We then deduce the following proposition. 

\begin{proposition} 
Suppose that $({\cal Z},X)$ and $({\cal Z}',X)$ are full twistor spaces 
for a generic linear system of ODEs on a Riemann surface $X$, with $G={\rm 
SL}(n,\C)$.  Let $a\in X$ be a pole of 
order $r+1>1$.  Then there are neighbourhoods $U,U'$ of $a$ in ${\cal Z}$ and 
${\cal Z'}$ and a ${\mathfrak g}$-equivariant biholomorphic map 
$\rho:U\to U'$ such that $\rho(a)=a$ and $\rho(X\cap U)= X\cap U'$. 
\label{locunique} \end{proposition} 

\begin{proof}  Choose a coordinate $z$ on a small disc in $X$ such that the 
pole is at $z=0$, and extend this to a neighbourhood of $a$ in ${\cal Z}$ so 
that $S$ is given by $z=0$.  Then we can identify a neighbourhood of $a$ with 
$S\times D$, as before. For each $s \in S$, we have a holomorphic map 
$\gamma_s:D\to {\cal Z}$ and hence an element 
$\alpha_s$ of ${\cal P}$ such that 
$\gamma_s^{{*}}=\alpha_s$. Let $\gamma'_s:D\to {\cal Z}'$ be the corresponding map 
into the second twistor space. Then the required biholomorphic 
map is $\rho:(s,z)\mapsto 
\gamma'_s(z)$. 
\end{proof}
Proposition \ref{globunique} above is an immediate corollary, since 
Proposition \ref{locunique} implies that the map $\rho$ constructed there 
extends to $S$.

\subsection*{Isomonodromic deformations}

We have shown that a generic ${\rm SL}(n,\C)$ 
system of the form (\ref{ODE}) on a Riemann surface can 
be generated from a twistor space $({\cal Z},X)$, and that, if we require 
${\cal Z}$ to be full, then it is unique, at least in a neighbourhood of 
$X$. In the case of Fuchsian equations on $\CP_1$, Hitchin (1995a) showed 
that the isomonodromic deformations of the system are given by the 
deformations of $X$ in ${\cal Z}$ (every twistor space being full in the 
Fuchsian case).  This is also true more generally, as we shall now see.

By Kodaira's theorem,  the deformations of a compact 
curve $X\subset {\cal Z}$ are determined by the properties of the normal bundle 
$N=T{\cal Z}\vert_{X}/TX$.  Put
$$
d_1=\di H^1(X,N)\qquad \mbox{and} \qquad d_0=\di H^0(X,N)\, .
$$
When $d_1=0$, the theorem implies that
$X$ is one of a complete $d_0$-parameter holomorphic family 
of embedded curves.  The tangent space to the parameter space at $X$ is 
naturally identified with $H^0(X,N)$ (Kodaira 1962).

For each curve $X$ in the family, we have a meromorphic 1-form 
$\theta\vert_X$ and hence a system of differential equations  of the form 
(\ref{ODE}). As we vary the curve along a path $X_t$ in the family, $t\in 
[0,1]$, the tangent at  $t$ is an element of $H^0(X_t,N)$. This we represent 
by local sections $Y_i$ of $T{\cal Z}\vert_{X_t}$, chosen to be tangent to $S$ 
at the poles. Thus the $Y_i$s are uniquely determined up to the addition local 
tangent vector fields to $X_t$ that vanish at $S\cap X_t$.  Put 
$\Omega_i=i_{Y_i}\theta$, $\alpha_t=\theta\vert_{X_t}$, 
and identify local neighbourhoods in the $X_t$s along $Y_i$. 
Then $\Omega_i$ is meromorphic, with a pole of order 
$r$ at a singularity of rank $r$. By (\ref{flatness}), 
$$
\p_t\alpha_t=\nabla_{\alpha_t}\Omega_i=\rd \, \Omega_i+[\alpha_t,\Omega_i]\, .
$$
Moreover on the overlap of their domains, $\Omega_i-\Omega_j=i^{\str} 
_{T_{ij}}\alpha_t\, \rd z$ for some tangent vector $T_{ij}$ to $X$, which must 
vanish at any poles in the overlap.  By using the results in the appendix,
we deduce the following proposition. 
\begin{proposition}
Let $G={\rm SL}(n,\C)$, let $({\cal Z},X)$ be a twistor space, and let $X'$ be 
a deformation of $X$.  Then the linear system of ordinary differential equations 
on $X'$ is an isomonodromic deformation of the linear system on $X$. 
\end{proposition}

\subsubsection*{The minimal twistor space}
In the minimal case, we can find the normal bundle of $X$ 
as follows.
For $x\in X$, put $L_{\alpha}(x)=\alpha(T_xX)
\subset {\mathfrak g}$ when $x$ is not a pole; and
$$L_{\alpha}(x)=z^{r+1}\alpha(T_xX)\, ,$$ 
when $x$ is a pole of rank $r$ and $z=0$ at $x$.
Then
$L_A\to X_0$ is a holomorphic line bundle.  
Moreover $\alpha$
is a global meromorphic section of $L_{\alpha}\otimes K$ with 
a pole of order $r+1$ at a singularity of rank $r$ and, by assumption, no 
zeros.  Therefore 
$$L_{\alpha}\otimes K=-
\sum 
(r_i+1)a_{\al}\, 
$$  
and so ${\rm deg}\, L_{\alpha}=2-2g-\sum (r_i+1)$, where the sum is over 
the singularities and $g$ is the genus. 

Now a point $x\in X$ is the image
of $(x,e)\in X\times G$ under the projection along $F$.
So we have 
$$
N_x=T_{(x,e)}(X\times G)/(T_x\oplus L_{\alpha})\, .
$$ 
Thus there is a short exact sequence
$$
0\to L_A\to {\mathfrak g}\to N \to 0\, ,
$$
where ${\mathfrak g}$ is the trivial bundle ${\mathfrak g}$-bundle over $X$. 

When $X=\CP_1$ and   $\sum (r_i+1)\geq 4$, we have
$H^1({\mathfrak 
g})=0$, $H^0(L_A)=0$, and 
$$
\di H^1(L_A)=-3+\sum (r_i+1)\, .
$$
From the corresponding long
exact sequence, therefore, $H^1(N)=0$, and
$$
0\to H^0({\mathfrak g})\to H^0(N)\to H^1(L_A)\to 0\, .
$$
It follows that
$$
d_1=0, \qquad d_0=\sum (r_{i}+1)-3+\di {G}\, ,
$$
We conclude that $X$ is one of a $d_0$-parameter of embedded copies of 
$\CP_1$. If all the singularities are regular, then $d_0$ is the 
dimension of the 
space of configurations of poles ($\sum(r_i+1)-3)$, plus the dimension of $G$.  
The deformations of $X$ are parametrized by the positions of the poles, modulo 
the action of ${\rm SL}(2,\C)$ on $\CP_1$, together with constant gauge 
transformations. 

\ex {\em The Schlesinger equations} Suppose that $X=\CP_1$ and that all the 
singularities are regular.  We choose the domains of the $\Omega_i$s so that 
each pole lies in only one. The $\Omega_i$s are holomorphic and the 
deformation is determined by the holomorphic tangent vector fields $T_{ij}$ on 
the overlaps of the domains. Since $H^1(X,TX)=0$, we have $T_{ij}=T_i-T_j$, 
where $T_i$ is holomorphic on the domain of $\Omega_i$ (but possibly non-zero 
at the corresponding pole).  So if we put $Y=Y_i+T_i$, then $Y$ is a global 
section of $T{\cal Z}\vert_{X}$.  

Let $z$ be a global stereographic coordinate on $X$, with $z=\infty$ 
not one of the poles, and use $Y$ to transfer
$z$ along the deformations of $X$.  Then
$$
\alpha=-\sum \frac{A_{i}\, \rd z}{z-a_i}
$$
where the $a_i$s are positions of the poles and the coefficients $A_i$ are 
independent of $z$ and satisfy
$$
\sum_i A_i=0
$$
(since there is no pole at infinity).  Put $\Omega=i_Y\theta$.  Then 
$\Omega$ is 
meromorphic with simple poles at the $a_i$s and 
$$
\frac{\p A}{\p t}=\frac{\p \Omega}{\p z}-[A,\Omega]\, .
$$
In fact, since $\Omega_i$ is 
holomorphic at $a_i$ and $T_i(a_i)=\p a_i/\p t$, we have
$$
\Omega = -\frac{\p a_i}{\p t}
\frac{A_i}{z-a_i}+O\bigl((z-a_i)^0\bigr)\, .
$$
It follows that
$$
\Omega = -\sum_i \frac{\p a_i}{\p t}
\frac{A_i}{z-a_i}+k
$$
where $k$ is a matrix independent of $z$. 
Therefore
$$
\frac{\p A_i}{\p t}=-\sum_{j\neq i}\frac{\p a_i}{\p 
t}\frac{[A_i,A_j]}{a_i-a_j}+[k,A_i]\, ,
$$
which is a form of the {\em Schlesinger equations}. The last term is simply a 
infinitesimal gauge transformation; the first gives the dependence of the 
$A_i$s on the configuration of the poles. 

\subsubsection*{The full case}
In a full twistor space, $d_0$ is generally larger than in the minimal construction.
Here we can find $N$ in another way.  

We suppose that all the singularities are generic.  Then, by (\ref{Omcond})
and Lemma \ref{lemm1}, a local section $V$ of $T{\cal Z}\vert_X$ is a map $\Omega$ from
a neighbourhood in $X$ to ${\mathfrak g}$ such that a singularity at of rank $r$, 
$$
\Omega=O(z^{-r}), \qquad [\alpha,\Omega]=O(z^{-r-1})\, ,
$$
where $z$ is a local coordinate that vanishes at the singularity. That is, in a 
local gauge in which the singular part of $\alpha$ is diagonal, the diagonal 
entries in $\Omega$ have poles of at most order $r$, and the off-diagonal entries
are holomorphic.  Thus these algebraic conditions charaterize $\Omega$ as a local 
section of a holomorphic bundle $E\to X$
with fibre 
${\mathfrak g}$ (and therefore rank $n^2-1$) and degree $\sum (n-1)r_i$. In the full case,
therefore, we have that $T{\cal Z}\vert X=E$ can be constructed directly from the positions
and ranks of the singularities of the ODE on $X$.

Put
$$
L=TX\otimes \sum (-a_i)\, ,
$$
so that a local holomorphic section of $L$ is a tangent vector field that 
vanishes at the poles.  Then we have a 
short exact sequence 
$$
0\to L\to E\to N\to 0\, 
$$
with the second map given by contraction with $\alpha$. Hence there is an exact sequence
$$
0\to H^0(L) \to H^0(E) \to H^0(N) \to H^1(L) \to H^1(E) \to H^1(N) \to 0\, .
$$
If the genus of $X$ is $g$, and if there are $k$ singularities 
in total, then
$$
{\rm deg}(L)=2-2g-k, \quad \di H^0(E)- \di H^1(E) 
=(n^2-1)(1-g)+(n-1)\sum r_i\, ,
$$
(the latter identity coming from the Riemann-Roch theorem). 

A global section 
of $E$ is a meromorphic map $\Omega: X\to {\mathfrak g}$ such that at a 
singularity of rank $r$, 
$$ 
\Omega=z^{-r}g^{\str}_{{\rm f}}qg_{{\rm f}}^{-1}+O(z^0)\,
$$
where $z$ is a local coordinate that vanishes at the singularity and $q$ is a 
diagonal polynomial of degree $r-1$.
When $X=\CP_1$, $\Omega$ is determined as a global rational map by 
the $q$s up to the addition of a constant element of ${\mathfrak g}$, and the $q$s 
can be specified independently.  In this case, therefore, $\di H^0(E)= 
(n-1)\sum r_i+\di G$, and $\di H^1(E)=0$.  Moreover if $k\geq 4$, then $$\di 
H^0(L)=0, \qquad \di H^1(L)=k-3\, . 
$$
It follows that
$$
d_0=H^0(N)=(n-1)\sum r_i+\di G+k-3, \qquad d_1=\di H^1(N)=0\, .
$$ 
If either $n=2$ ($G={\rm SL(2,\C)}$) or $\sum r_i=0$ (all singularities 
regular), then $d_0$ is the same as in the minimal case; in either of these 
cases, the minimal twistor space is full and, by Proposition \ref{allisoprop} 
below, it gives all possible isomonodromic deformations.  In general, however, 
there are more isomonodromic deformations than are given by the minimal 
construction: the additional parameters are the coefficients of the diagonal 
polynomials $t$ (of degree $r-1$) at the irregular singularities ($r\geq 1$). 

When $X$ has higher genus, $\di H^1(E)=\di H^0(E^{*}\otimes K)^{*}$ 
is generically zero whenever $(n-1)\sum r_i>n^2g+g-2$.

\subsubsection*{Twistor curves}

Let $\rd y+\alpha y=0 $ be a generic system on a compact Riemann surface $X$ 
and suppose that $H^1(E)=0$.  Then we can construct a full twistor space 
$({\cal Z},X)$. Since $H^1(X,N)=0$, $X$ is one of a complete holomorphic 
family ${\cal K}$ of curves $X\subset {\cal Z}$. 

\begin{proposition} Let $(X_t,\alpha_t)$, $t\in [0,1]$, be an isomonodromic 
deformation of $(X,\alpha )$.  Then for small $t$, there is a path 
$X_t$ in ${\cal K}$ such that 
${\alpha}_t={\theta}\vert_{X_t}$.
\label{allisoprop} \end{proposition} 
\begin{proof} Let $y_t$ be  solution to 
$$
\rd y_t+\theta_ty_t=0
$$
with constant monodromy, and with constant connection matrices $C_{i}$ to 
the special solutions at the poles.  

Let $z,z'\in X$ be nearby points (neither a pole) and let $g,g'\in G$ be close 
to the identity.  Then, by integrating the action of ${\mathfrak g}$ on ${\cal 
Z}$, we have two points $zg$, $z'g'$ in ${\cal Z}$ near $X$.  These are the 
same if 
\be
gg^{\prime -1}=y_0(z)y_0(z')^{-1}\, .
\label{ggprime}\ee

Let $z_t\in X_t$ vary continuously with $t$, and suppose that $z_t$ is not a 
pole for any small $t$.  Put 
\be
\rho_t(z_t)=z_0y_0(z_0)y_t(z_t)^{-1}\in {\cal Z}
\label{rhodef}\ee
(the right-hand side is interpreted by regarding $z_0\in X$ as a point of 
$X\subset {\cal Z}$ and by using the local action of $G$ on ${\cal Z}$).  This is 
independent of the choice of branch of $y_t$ and $y_0$ (so long as we make the 
choice of branch continuously) since $y_0$ and $y_t$ have the same monodromy.  
Moreover, $\rho_t(z_t)$ depends only on $z_t$, and not on the path, by 
(\ref{ggprime}).  So if we exclude a small neighbourhood of each pole in $X_t$, 
then we can embed the complement in ${\cal Z}$ by $z_t\mapsto \rho_t(z_t)$.  
By fixing $z_0$ and moving $z_t$, we see from (\ref{rhodef}) that 
$\rho_t^{{*}}\theta=\alpha_t$.

It remains to show that $\rho_t$ extends holomorphically to the poles. 
Consider one of the poles (a point of $X_t$, varying continuously with $t$).  
We can choose a 
coordinate $z$ in a neighbourhood $D$ of the pole on each $X_t$ so that $D$ is 
the unit disc and the pole is at $z=0$.  Then, for small $t$, since ${\cal Z}$ 
is full, there exists a holomorphic map $\gamma_t:D\to {\cal Z}$ such that 
$\alpha'_t=\gamma_t^{*}\theta$ has the same singularity data at $z=0$ as 
$\alpha_t$ has at $z=0$.  Since $\alpha_t$ is an isomonodromic deformation, 
$\alpha_t$ and $\alpha'_t$ also have the same Stokes' matrices.  

Let $y'_t$ be a solution to 
$$
\rd y'_t+\alpha'_t=0
$$
with the same monodromy and connection matrices to the special solutions in 
the sectors at $z=0$ as $y_0$.  Then 
$$
\gamma_t(z)=zy_0(z)y'_t(z)^{-1}\, .
$$
Further $y'_ty_t^{-1}$ is 
holomorphic at $z=0$.  This is because it is single-valued, since
$y_t$ and $y'_t$ have the same holonomy, and bounded since in any sector
${\cal S}$ at $z=0$
$$
y'_ty_t^{-1}= y'_{\cal S}y_{\cal S}^{-1}
            \sim g'_{{\rm f}}g_{{\rm f}}^{-1}
$$
where $y'_{\cal S}$, $y^{\str}_{\cal S}$ 
are the corresponding special solutions and 
$g'_{{\rm f}}$ , $g^{\str}_{{\rm f}}$ are the formal gauge transformations to 
diagonal form.
So the embedding $\rho_t$
extends by 
mapping $z\in D\subset X_t$ to $\gamma_t(z)y'_ty_t^{-1}$. 
\end{proof}

\subsubsection*{Isomonodromic flows for systems on the Riemann sphere}
The number of independent 
isomonodromic deformations (the dimension of ${\cal K}$) 
of a generic system on $X=\CP_1$, 
$$
\di H^0(X,N)=(n-1)\sum r_i+\di G+k-3\, .
$$
We shall now show that the deformations 
are given by Hamiltonian flows on symplectic 
manifolds constructed from the affine orbits in $L{\mathfrak g}^{*}$.

In this case, 
the twistor curves in ${\cal Z}$ are copies of $\CP_1$, and they 
can be 
parametrized by a global stereographic coordinate $z\in \C\cup \{\infty\}$.
We denote  by $\hat {\cal K}$ the space of parametrized curves, which has 
dimension
$$
\di \hat {\cal K}=
(n-1)\sum r_i+\di G+k\, .$$ 
The points of $\hat {\cal K}$ can be labelled 
by the positions of the poles ($k$ parameters), the polynomials $t$ at each 
pole ($(n-1)\sum r_i$ parameters) and a choice of gauge ($\di G$ parameters).

An element of $\hat {\cal K}$ is a mapping $\rho:\CP_1\to {\cal Z}$ from some 
fixed copy of the Riemann sphere.  It determines a rational ${\mathfrak g}$-valued 
$1$-form 
$$
\alpha=\rho^{{*}}\theta=-A\, \rd z\, ,
$$
where $A$ is rational, with poles of order $r_i+1$ at $k$ points $a_1, \ldots 
,a_k$ (we assume that
none of the poles is at infinity, so $A=O(z^{-2})$ as $z\to 
\infty$). In a neighbourhood of $a_i$, we put $z_i=z-a_i$ and assume, 
without of loss of generality, that no other pole lies in the closure of
the disc $D_i=\{\vert z_i\vert <1\}$. Then  
$\alpha_i=\alpha\vert_{D_i}$
determines a point of the symplectic 
manifold ${\cal M}_{r_i}$. Thus we have a map
$$
\hat {\cal K}\to
{\cal M}={\cal M}_{r_1}\times {\cal M}_{r_2}\times 
\cdots \times {\cal M}_{r_k}\, .
$$
It is not surjective, since a given point of ${\cal M}$ is not, in 
general, given by the restrictions of a global 1-form $\alpha$.  However 
$\alpha$ is uniquely determined by the positions of its poles and by its 
image in ${\cal M}$, since the difference between two $\alpha$s with 
the same pole positions, and determining the same point of ${\cal M}$, 
is a global holomorphic $1$-form, and therefore vanishes. 

Given $[\alpha_i]\in {\cal M}_{r_i}$ and the points $a_i\in \CP_1$, 
we put $\alpha_i=-A_i\, \rd z_i$ 
and denote by $A_{i-}$ and $A_{i+}$ 
the negative and non-negative degree terms in the Laurent expansion of 
$A_i$ in powers of $z_i$ in a neighbourhood of $a_i$.   Given also 
a diagonal polynomial $q_i$ of degree $r_i-1$, we put
$$
\Omega_{q_i}=\Bigl(z^{-r_i}g_{{\rm f}}^{-1}q_ig^{\str}_{{\rm f}}\Bigr)_{-}
$$
where $g_{{\rm f}}$ is the formal gauge transformation to the diagonal form of 
$\alpha_i$ and again the minus subscript
denotes the negative terms in the Laurent 
expansion in powers of $z_i$. Then $A_{i-}$ and $\Omega_{q_i}$ are 
global meromorphic functions on $\CP_1$ with values in ${\mathfrak g}$; they are
holomorphic except at $a_i$, where they have poles of order $r_i+1$ and $r_i$, 
respectively. Moreover $A_{i-}$ and $\Omega_i$ are independent of the choice 
of representative in $[\alpha_i]$.

\begin{proposition} The isomonodromic deformations of a generic ${\rm 
SL}(n,\C)$ system on $\CP_1$ are generated by the 
Hamiltonians
$$
h_v=\sum_j
\frac{1}{2\pi\ri}\oint \tr(\alpha_j v), ~~~ 
h_i=\sum_{j\neq i}\frac{1}{2\pi\ri}\oint \tr(\alpha_j A_{i-}), ~~~ 
h_{q_i}=\sum_j\frac{1}{2\pi\ri}\oint\tr \bigl(\alpha_j\Omega_{q_i}\bigr)\, ,
$$
on ${\cal M}$, where 
$v$ is a constant element of ${\mathfrak g}$
and the integrals are around small circles surrounding the poles.
\end{proposition}
The $h_i$s are time-dependent Hamiltonians, the `times' being the positions $a_i$ of the 
poles.

\begin{proof}  First we note that the Hamiltonians 
$h_v$ generate the constant gauge 
transformations.
Consider next the flow generated by $h_i$. We shall find the 
value of the Hamiltonian vector field at a point of ${\cal M}$ 
constructed from a global meromorphic 1-form $\alpha$.  To do this, we must 
we must evaluate the 
gradient of $h_i$ at such a point. We have
\begin{eqnarray*}
\delta h_i&=&\sum_{j\neq i}\frac{1}{2\pi\ri}\oint \tr(\delta \alpha_j A_{i-}+
\alpha \,\delta A_{i-})\\
&=& -\frac{1}{2\pi\ri}\oint \tr(\alpha_i \,\delta A_{i-})+
\sum_{j\neq i}\frac{1}{2\pi\ri}\oint \tr(\delta \alpha_j A_{i-})\, .\\
\end{eqnarray*}
However
\begin{eqnarray*}
\frac{1}{2\pi\ri}\oint \tr(\alpha_i \,\delta A_{i-})&=&
\frac{1}{2\pi\ri}\oint \tr(\alpha_{i+} \,\delta A_{i})\\&=&
\frac{1}{2\pi\ri}\oint \tr(A_{i+} \,\delta \alpha_{i})\, .\\
\end{eqnarray*}
We conclude that the value of the Hamiltonian vector field at such a point is 
$$
\delta \alpha_i=-\nabla A_{i+}, \qquad 
\delta \alpha_j=\nabla A_{i-} \quad j\neq i \, .
$$

The claim is that this is tangent to an 
isomonodromic deformation.  To see this, let $y$ 
be a solution to $\rd y+\alpha y=0$, let $D$ be a disc containing $a_i$, but 
no other pole, and let $D'$ be a second disc not containing $a_i$ such that 
$D,D'$ is an open cover of $\CP_1$. For small $t$, put 
$F_t(z)=
y(z-t)y(z)^{-1}$. Then $F:D\cap D'\to G$ is single-valued, holomorphic, and 
equal to the identity when $t=0$.  By Birkhoff's theorem,
$F_t=h_t^{-1}h'_t$ for some holomorphic maps $h_t:D\to G$, $h'_t:D'\to G$, 
with $h'_t(\infty)=1$.

Put 
$$
y_t(z)=\left\{\begin{array}{ll} h_t(z)y(z-t)& z\in D\\h'_t(z)y(t)& z\in D'
\end{array}\right.
$$
and $\alpha_t=-\rd y_t\, y_t^{-1}$.  Then the definitions agree on $D\cap D'$ 
and $\alpha_t$ is a global meromorphic 1-form with poles at $z=a_j$ ($j\neq 
i$) and $z=a_i+t$.  Moreover $\p_ty_t=\Omega_ty_t$, where
$$
\Omega_t=\left\{\begin{array}{ll} 
\p_t h^{\str}_t h_t^{-1}-h_t A_i(z-t)h_t^{-1}& \mbox{in $D$}\\
\p_t h'_t h_t^{\prime -1}& \mbox{in $D'$.}
\end{array}\right.
$$
Since $h_t$ and $h'_t$ are holomorphic in $D$ and $D'$,
it follows that the deformation is isomonodromic (see Proposition \ref{isochar}).

At $t=0$, we have $h_t=h'_t=1$ and
$$
\p_t h'_t =A_{i-}\qquad \p_th_t=-A_{i+}\, ;
$$
we also have at $t=0$ that $\delta \alpha_i=\nabla(\p_t h_t)$,
$\:\delta \alpha_j=\nabla(\p_t h'_t)$ for $j\neq i$.  So the tangent to the 
deformation is the Hamiltonian vector field constructed above: these 
deformations move the poles, but leave the singularity data unchanged.

Now consider the flow generated by $h_{q_i}$.  Proceeding as before to 
calculate the value of the Hamiltonian vector field at a point given by a 
global 1-form $\alpha$, we have
\begin{eqnarray*}
\delta h_{q_i}
&=&
\sum_j\frac{1}{2\pi\ri}\oint\tr \bigl(\delta 
\alpha_j\Omega_{q_i}+\alpha_j\delta \Omega_{q_i}\bigr)\\
&=&
\sum_j\frac{1}{2\pi\ri}\oint\tr \bigl(\alpha_j\delta \Omega_{q_i}\bigr)\, .
\end{eqnarray*}
So in this case, the value of the Hamiltonian vector field is
$$
\delta \alpha_i=\nabla_{\alpha}\Omega_{q_i}\, ,
$$
which is clearly isomonodromic.  These deformations change the singularity 
data at $a_i$, leaving the position of the poles unchanged. \end{proof} 

\begin{remark}  The Hamiltonians $H_{q_i}$ vanish on the orbit of a global 
1-form $\alpha$, while
$$
h_i=\frac{1}{\tau}\frac{\p\tau}{\p a_i}\, ,
$$
where $\tau$ is as in Jimbo {\em et al} (1981a).
\end{remark}

\ex  In the case of a generic system with only regular singularities,
${\cal M}={\cal M}_1\times \cdots \times {\cal M}_1$.  The tangent space is 
spanned ${\cal M}_1$ is spanned by the generators of constant gauge 
transformations.  If we write
$$
\alpha= -A\, \rd z= \sum \frac{A_i\, \rd z}{z-a_i}, \qquad \alpha_i 
=-\frac{A_i\, \rd z_i}{z_i}\, ,
$$
then we can identify ${\cal M}$ (as a symplectic manifold) with the product of 
the coadjoint orbits of the $A_i$ in ${\mathfrak g}^{{*}}$.  In this case, 
$$
A_{i-}=\frac{A_i}{z-a_i}\, ;
$$
there 
are no $h_{q_i}$s, while
$$
h_i=\sum_{j\neq i}\frac{\tr(A_iA_j)}{a_i-a_j}\, .
$$

\section*{Appendix}

\subsection*{Singularities of systems of ODEs}

Let $A$ be a meromorphic map from some neighbourhood $D$ of the origin in $\C$ 
into ${\mathfrak g}$ (the Lie algebra ${\rm sl}(n,\C)$), with a pole of order $r+1$ 
at $z=0$.  Then the system
\be
\frac{\rd y}{\rd z}=Ay\, ,
\label{locODE}\ee
has a {\em singularity} of {\em Poincar\'e rank} $r$ at the origin. It is 
{\em regular} or {\em Fuchsian} if $r=0$, and {\em irregular} if $r>0$.
In this paper, $y$ will always be a {\em matrix fundamental solution}---that 
is, it will take values in $G$.

\subsubsection*{Gauge and point transformations}

When we regard the system as a connection on a vector bundle,
we must allow {\em gauge transformations} (changes of trivialization) 
of the form
$$
A\mapsto g^{-1}Ag -g^{1}\frac{\rd g}{\rd z}, \qquad \mbox \qquad \nabla\mapsto 
g^{-1}\circ \nabla\circ g,\qquad y\mapsto gy\, ,
$$
where $g:X\to G$ is holomorphic. When $g$ is constant, $A$ transforms by 
conjugation. We also admit transformations of the coordinate $z\mapsto \hat z$, 
under which
$$
A\mapsto \hat A=A\frac{\rd z}{\rd \hat z}\, .
$$
If this is to fix the singularity at the origin, then we must take 
$\hat z(0)=0$.

\subsubsection*{Generic irregular singularities}

An irregular singularity of rank $r$ is {\em generic} if the eigenvalues of
$$
A_{-r-1}=z^{r+1}A\vert_{z=0}
$$
are distinct. In this case, we can assume, without loss of generality, that 
the eigenvalues of $z^{r+1}A$ are distinct throughout the neighbourhood. 
If we choose an ordering for the eigenvalues, then we can find a holomorphic 
map $g:D\to G$ such that $z^{r+1}g(z)A(z)g^{-1}(z)$ is holomorphic and 
diagonal, with the eigenvalues as diagonal entries. 

It follows that we can find a holomorphic map $D\to G$ and a diagonal 
polynomial $p(z)$ of degree $r$ such that
$$
A-g(z)^{-1}\frac {p(z)}{z^{r+1}}g(z) 
$$
is holomorphic at the origin.  Therefore we have the 
normal form:
\be
A\sim \rd\left(\frac{t}{z^r}\right)+\frac{m\, \rd 
z}{z}+R(z)\, 
\label{normform1}\ee
where `$\sim$' denotes gauge equivalence, $t$ is a diagonal polynomial of 
degree $r-1$, $m$ is a constant diagonal matrix called the {\em exponent of 
formal monodromy}, and the remainder $R$ is holomorphic at $z=0$. 
Given the local coordinate $z$ and the ordering of the eigenvalues, $t$ and $m$ 
are uniquely determined by $A$, independently of the choice of gauge.  We call 
them {\em the singularity data} at $0$.  If the 
ordering is changed, then the diagonal entries are permuted; if the coordinate 
is changed, then $m$ is unchanged, while $t\mapsto t'$, where $t'$ is obtained 
from $t$ by making the coordinate transformation and truncating the Taylor 
series in $z$.     

If one looks for a gauge transformation $g(z)=g_0+g_1z+\cdots $ such that 
(\ref{normform1}) holds with $R=0$, then the coefficients $g_i$ can be 
determined uniquely once a choice has been made for $g_0$ to diagonalize 
$A_{-r-1}$.  For each such choice, one can therefore find a unique 
formal solution
$$
y^{\str}_{{\rm f}}=g^{\str}_{{\rm f}}(z)\exp\bigl(Tz^{-r}+m\log z\bigr)\, .
$$
In general, the formal series does not converge.  However, by truncating, one 
can make $R$ vanish to arbitrarily high order at $z=0$.

\subsubsection*{Sectors and special solutions}

The eigenvalues $\lambda_i$ of $A_{-r-1}$ determine a sequence of {\em 
Stokes' rays}  through the origin, on which 
${\rm Re}\bigl(z^{-r}(\lambda_i-\lambda_j)\bigr)$ changes sign for some pair 
of eigenvalues. Given a pair of consecutive rays (in order around the unit 
circle), with arguments $\theta_1,\theta_2$, we define a {\em sector} ${\cal S}$ by
\be
{\cal S}=\{z\:\vert\: \theta_1-\pi/2r<{\rm arg}(z)<\theta_2+\pi/2r\}\, .
\label{sect}\ee 
For each such sector ${\cal S}$, there is a unique {\em special solution}
$$
y^{\str}_{\cal S}=g^{\str}_{\cal S}(z)\exp\bigl(tz^{-r}+m\log z\bigr)\, ,
$$
such that $g^{\str}_{\cal S}\sim g_{{\rm f}}$ as $z\to 0$ in ${\cal S}$.  
(See, for example, 
Boalch 2000 for a careful account of this proposition).

The solutions $y^{\str}_{\cal S}$ are independent of the choice of coordinate (as maps $D\to 
G$).  They are uniquely determined in each sector by the choice of $g_0$. 
So, as sections of the principal bundle over each sector, 
they are determined uniquely by the choice of the frame at the origin in which 
$A_{-r-1}$ is diagonal.

\subsubsection*{Regular singularities}
In the regular case ($r=0$), (\ref{normform1}) still holds, with $t=0$,
provided that the eigenvalues of $A_{-1}$ are distinct.  The formal series for
$g$ can be found provided that, in addition, no two eigenvalues differ by an 
integer.  It then necessarily converges, so $y_{{\rm f}}$ is a solution.

\subsubsection*{Global systems}

Let $X$ be a compact Riemann surface. For each $k$-tuple ${\bf r}= 
(r_1,\ldots, r_k)$ of nonnegative integers, we denote by ${\cal D}_{{\bf 
r}}(X,E)$, or simply by ${\cal D}_{{\bf r}}$, the space of meromorphic ${\rm 
sl}(n,\C)$-connections $\nabla=\rd +\alpha$ 
on the trivial vector bundle $E=X\times \C^n$ 
with $k$ poles of order (at most) $r_1+1, \ldots, r_k+1$.\footnote{Note that 
$\alpha$ determines a holomorphic map $X\to {\mathbb P}{\mathfrak g}$}

If we choose a local trivialization and 
a coordinate $z$ that vanishes at one of the poles,
a connection $\nabla\in {\cal D}_{{\bf r}}$ is given 
in a neighbourhood of the pole by (\ref{locODE}).\footnote{ When $X=\CP_1$, 
we have $\alpha=-A\rd z$, where $z$ is a stereographic coordinate and $A$  is 
defined globally as a rational section of ${\mathfrak g}\otimes {\cal O}(-2)$. If 
all its singularities are at finite values of $z$, then $A$ has $N$ poles and 
$A=O(z^{-2})$ as $z\to \infty$; but if one of the singularities is at 
infinity, then $A=O(z^{r-1})$ as $z\to \infty$, where $r$ is the rank.} 

\subsubsection*{The monodromy representation}

A local $G$-valued solution $y$ to the equation $\rd +\alpha y=0$
can be continued analytically: 
it is singular at the poles, and multi-valued 
(single-valued on the covering space of the complement of the poles). 
If 
$$
\gamma:[0,1]\to\CP_1\setminus \{a_1, \ldots ,a_k\}, \qquad 
\gamma(0)=\gamma(1)=z_0,
$$
is a closed loop and $z_0$ is some fixed base point, then 
we have $y(\gamma(1))=M_\gamma\in G$ for some {\em monodromy matrix} 
$M_{\gamma}$ which depends only on the homotopy class of $\gamma$. 
\begin{definition} 
The {\em monodromy representation} of {\em (\ref{ODE})} is the 
homomorphism
$$
\pi_1\bigr(\CP_1\setminus \{a_1, \ldots ,a_k\}\bigl) \to G: [\gamma]\mapsto
M_{\gamma}\, .
$$
\end{definition}
The monodromy representation is independent of $z_0$ and the choice of $y$
up to conjugation by a fixed element of $G$.

\subsubsection*{Deformations}

\begin{definition} Let $\nabla_0,\nabla_1
\in {\cal D}_{{\bf r}}(X)$. A {\em deformation} 
of $\nabla_0$ into $\nabla_1$ is a 
smooth path $\nabla_t\in {\cal D}_{{\bf r}}(X)$, $t\in 
[0,1]$, from $\nabla_0$ to $\nabla_1$. \end{definition} 
We are interested in the deformations of a connection $\nabla_0$ into 
a second $\nabla_1$ (or equivalently of $\alpha_0$ into $\alpha_1$) 
while preserving certain 
properties of the corresponding linear system. 
\begin{proposition} A deformation $\nabla_t$ has 
constant monodromy representation 
(up to conjugation) if and only if
$$\frac{\rd \nabla_t}{\rd t}=\nabla_t\Omega_t
$$
for some family of holomorphic maps $\Omega_t$ (depending smoothly on $t$) 
from the complement of the poles of $\nabla_t$ in $X$ into ${\mathfrak g}$.
\end{proposition}
There is an awkwardness in the terminology here: 
it is important to keep in mind that `having the same monodromy 
representation' is not the same as `isomonodromic' when irregular 
singularities are present.

\begin{proof} By fixing a base point (disjoint from the poles)
and a frame at the base point, we can find a solution $y_t$ for each $t$ which 
depends smoothly on $t$.  If the monodromy representation is constant, then we 
can find a matrix $K_t\in G$  for each $t$ such that the monodromy matrices 
of $y_tK_t$ are constant.  If we take $t$ close to $t'$ and exclude small 
discs around the poles of $\nabla_t$, then
$g_{tt'}=y_tK_tK_{t'}^{-1}y_{t'}^{-1}$ is single-valued, 
and we can 
construct $\Omega_t$ in (C1) by putting
$$
\Omega_t= \left.\frac{\p g_{tt'}}{\p t'}\right\vert_{t'=t}\, .
$$
(This is holomorphic except at the poles of $\nabla_t$).

Conversely, if we are given $\Omega_t$, then $\rd - A\, \rd z -\Omega\,\rd t$ 
is a flat connection on the trivial bundle over 
$$
(X\setminus \mbox{poles})  \times [0,1]\, .
$$  
Its holonomy coincides with the monodromy $A_t$ for each $t$, and so the 
monodromy representation must be constant up conjugacy. 
\end{proof}
\subsubsection*{Isomonodromic deformations}
We now consider deformations 
$(X_t,\nabla_t)$ in which we change both $X$ and the 
connection $\nabla$.
 
Let $a$ be a pole of $\nabla$ and ${\cal S}$ a sector at $a$.  Then we have a solution 
$y^{\str}_{{\cal S},a}$ to the system $\nabla y=0$, uniquely determined up to the choice of the 
frame at $a$ in which the leading coefficient of $A$ is diagonal. If $a'$ and 
${\cal S}'$ are another pole and sector at $a'$, and if $\gamma$ is a path 
with initial point near $a$ in ${\cal S}$ and endpoint near $a'$ in ${\cal S}'$, then we can 
continue $y^{\str}_{{\cal S},a}$ along $\gamma$.  We shall have
$$
y^{\str}_{{\cal S},a}=y^{\str}_{{\cal S}',a'}
C^{\str}_{{\cal S},{\cal S}',a,a',\gamma}
$$
where $C^{\str}_{{\cal S},{\cal S}',a,a',\gamma}$ is a constant matrix. These matrices are uniquely 
determined by $A$ up to
$$
C^{\str}_{{\cal S},{\cal S}',a,a',\gamma}\mapsto D^{\str}_{a}
C^{\str}_{{\cal S},{\cal S}',a,a',\gamma}D_{a'}^{-1}\, ,
$$
where, for each pole $a$, $D_{a}$ is the product of a diagonal matrix and a 
permutation matrix. The matrices connecting the special solutions in adjacent 
sectors at the same pole are called {\em Stokes' matrices}.

As we deform $\nabla$ and $X$, we can vary 
${\cal S},{\cal S}',\gamma$, 
and the special solutions continuously.
\begin{definition}
A deformation is {\em isomonodromic} if the exponents of formal monodromy and 
the matrices $C^{\str}_{{\cal S},{\cal S}',a,a',\gamma}$ 
are constant, for an appropriate choice of 
special solutions.
\end{definition}
An isomonodromic deformation has constant monodromy representation, but the 
converse is not true except in the Fuchsian case (all singularities regular). 
The following characterization of the isomonodromy property is implicit in
Jimbo, Miwa, and Ueno (1981a).

We can cover $X_t$ by discs $D_i$ varying continuously with $t$ such that each 
pole lies in just one disc.  On each disc $D_i$, we can choose a coordinate 
$z_i$ 
such that, if $a_i\in D$ is a pole, then $z_i=0$ at $a_i$, independently of $t$.
We shall use these coordinates to identify the discs as $t$ varies.
\begin{proposition} A deformation $(X_t,\nabla_t)$ with constant monodromy 
representation is isomonodromic if and only if 
\begin{quote}
(i) in $D_i$, $\rd \nabla_t/\rd t=\nabla_t\Omega_{it}$ for some meromorphic 
$\Omega_{it}: D_i\to {\mathfrak g}$ (depending smoothly on $t$); 
$\Omega_{it}$ is holomorphic except possibly at a 
singularity of $\alpha_t$, where it has a  pole of order at most $r_i$ if 
$\alpha_t$ has a singularity of rank $r_i$ in $D_i$;

(ii) on the intersection $D_i\cap D_j$ of two discs
$\Omega_{it}-\Omega_{jt}=i_{T_{ij}}\alpha_t$ for some holomorphic vector field 
$T_{ij}$.
\end{quote}
\label{isochar}
\end{proposition}
\begin{proof} We shall look at the proof in outline. Suppose that the 
deformation is isomonodromic.  Let $y_t$ be a solution to $\nabla_t y_t=0$, 
depending continuously on $t$ and with constant monodromy (we have to keep in 
mind that $y_t$ is multi-valued and singular at the poles). 

Let $D_i$ be a disc containing a pole (at $z_i=0$)
and put 
$$
g_{itt'}(z_i)=y^{\str}_{{\cal S}}(z_i)y^{\prime -1}_{{\cal S}'}(z_i)
$$
where $y^{\str}_{{\cal S}}$ and $y'_{{\cal S'}}$ 
are the special solutions at $t$ and $t'$ in the corresponding 
sectors at one of the poles.  Then, for $t'$ close to $t$,  $g_{tt'}$ is a 
single-valued holomorphic map $D_i\setminus \{0\}\to {\mathfrak g}$; it is 
independent of sector, because the Stokes' matrices are the same at $t$ and 
$t'$. Once it is established that it is possible to differentiate the 
asymptotic expansions term-by-term, it is immediate that 
$\Omega_{it}=\p_{t'}g_{tt'}\vert_{t'=y}$ is meromorphic, and of the required 
form. 

On a disc $D_i$ that does not contain a singularity, we put 
$g_{itt'}=y_t(z_i)y_{t'}(z_i)^{-1}$, and define $\Omega_{it}$ in the same way. 
If we choose the branch of $y_t$ to vary continuously with $t$, 
$g_{itt'}$ is independent of the choice of branch because the monodromy of 
$y_t$ is independent of $t$.  

Given $y_t$, the only freedom in the construction of $\Omega_{it}$ is in the 
choice of the coordinate $z_i$, and hence in the local identification of the 
discs on the different Riemann surfaces. A different choice for each $t$ will 
add $i_{T}\alpha$ to $\Omega_{it}$ for some local holomorphic vector field 
$T$.  Thus (iii) holds on the overlap of two discs.

To prove the converse, suppose that $\Omega_{it}$ is meromorphic, as stated. 
Choose a continuously varying sector ${\cal S}$ at the pole, and let 
$$
y^{\str}_{{\cal S}}(z_i,t)=g^{\str}_{{\cal S},a}(z_i)\exp\bigl(tz_i^{-r}+m\log z_i\bigr)\, ,
$$ be the 
corresponding special solution. Then, by writing $\nabla_t=\rd_z-A_t\, \rd z_i$ 
and dropping the subscripts, 
we have
$$
\frac{\p}{\p z}\left(\frac{\p y}{\p t}-\Omega y\right)=
\frac{\p A}{\p t}y+A\frac{\p y}{\p t}-\frac{\p \Omega}{\p t}y-\Omega Ay
=A\left(\frac{\p y}{\p t}-\Omega y\right)\, .
$$
It follows that 
$$
\frac{\p y}{\p t}-\Omega y=yK
$$
for some matrix $K$, which can depend of $t$ but not $z$. Therefore
$$
g_{{\cal S}}^{-1}\left[\frac{\p g^{\str}_{{\cal S}}}{\p t}+g^{\str}_{{\cal S}}\frac{\p }{\p 
t}\left(\frac{t}{z^r}\right)-\Omega g^{\str}_{{\cal S}}\right]
 =
\re^{tz^{-r}+m\log z}K
\re^{-tz^{-r}-m\log z}\, .
$$
The left-hand side is asymptotic to a power series, divided by $z^{r+1}$, 
as $z\to 0$ in ${\cal S}$ (the same series for each sector at the pole).
In the case $r>0$, 
each off-diagonal entry on the right-hand side has an exponential factor 
which must blow up as $z\to 0$ along some directions in ${\cal S}$ 
since the angle of the sector ${\cal S}$ is more than $\pi/r$. This is a 
contradiction unless the off-diagonal entries in $K$ all vanish. Thus $K$ is a 
$z$-independent diagonal matrix.  It can be absorbed into the special solutions 
to give that 
$$
\frac{\p y^{\str}_{{\cal S}}}{\p t}=\Omega_ty^{\str}_{{\cal S}}
$$
and hence that the $C$ matrices are constant. This is also true, more simply, 
in the regular case since the formal solutions then converge.
\end{proof}

\subsection*{Symplectic form on ${\cal C}_r$}

We prove here that the tensor in (\ref{omdef}) is a symplectic form on ${\cal C}_r$.

\begin{proposition} $\omega$ is a symplectic form on ${\cal C}_r$, independent 
of the choice of 
$C_1$.
\end{proposition}
\begin{proof}
From the definitions,
$C_{2r+1}=\re^{-2\pi\ri m}C_1M$ and, in variational notation,
\be\label{SC}
\delta S_i\, S_i^{-1}=C^{\str}_i\bigl( 
C_i^{-1}\, \delta C^{\str}_i
-C_{i+1}^{-1}\, \delta C^{\str}_{i+1}\bigl)C^{-1}_i\, .
\ee                                   
We must show that $\omega$ is skew-symmetric, closed, and non-degenerate. From the first constraint, we have 
$\tr(\delta S^{\str}_iS_i^{-1}\delta' S^{\str}_i S^{-1}_i)=0$. It follows that
\begin{eqnarray*}
0&=&\sum_1^{2r}\tr\Bigl(\delta S^{\str}_iS_i^{-1}\delta' S^{\str}_i 
S^{-1}_i\Bigr)\\
&=&\sum_1^{2r}\tr\Bigl(
(C_i^{-1}\delta C^{\str}_i-C_{i+1}^{-1}\delta C^{\str}_{i+1})
(C_i^{-1}\delta' C^{\str}_i-C_{i+1}^{-1}\delta' C^{\str}_{i+1})\, .
\end{eqnarray*}
However
\begin{eqnarray*}
\sum_1^{2r}\tr\bigl(C_{i+1}^{-1}\delta C^{\str}_{i+1}
C_{i+1}^{-1}\delta' C^{\str}_{i+1}\bigr)
&=&\sum_1^{2r}\tr\bigl(C_{i}^{-1}\delta C^{\str}_{i}
C_{i}^{-1}\delta' C^{\str}_{i}\bigr)
-4\pi^2\tr(\delta m\delta' m)\\
&&\quad {}-2\pi\ri\, \tr\bigl(
\delta m \delta' C^{\str}_1C_1^{-1}
+\delta' m \delta C^{\str}_1C_1^{-1}\bigr)\, .
\end{eqnarray*}
The skew-symmetry follows. A similar calculation, starting from 
$$
\tr(\delta S^{\str}_iS_i^{-1}\delta' S^{\str}_iS_i^{-1}\delta'' S^{\str}_iS_i^{-1})=0\, ,
$$ 
shows that $\omega$ is closed.
To show that $\omega$ is nondegenerate, we note that if $\omega(Y, \:\cdot\:)=$, then
$\delta C^{\str}_1C_1^{-1}-\pi\ri \delta m$ is anti-diagonal, 
$P^{\str}_i\delta C^{\str}_iC_iP_i^{-1}$ is lower triangular for each $i$, and 
$P^{\str}_i\delta C_i^{\str}C_i^{-1}P_i^{-1}$ is upper triangular. However, from (\ref{SC}),
$$
\delta C^{\str}_{i+1}C_{i+1}^{-1}
=S^{-1}_i\delta C^{\str}_iC_i^{-1}S_i -S_i^{-1}\delta C^{\str}_iC_i^{-1}S_i\, .
$$
Therefore $\delta C^{\str}_iC_i^{-1}$ is diagonal and so
$\delta C^{\str}_iC_i^{-1}=\pi\ri\delta m$ for each $i$. It then follows from the second constraint (C2)
in the defintion of
${\cal C}_r$
that $\delta m=0$.

If we make a different choice for $C_1$ at each point, then the effect is to 
replace $C_i$ by $C_iK$, where $K$ is independent of $i$.  This adds
$$
\frac{1}{2\pi\ri}\sum_1^{2r} \tr\Bigl(C_i^{\str}\delta K C_i^{-1}\delta' 
S^{\str}_iS_i^{-1}\Bigr)-\tr(C^{\str}_1\delta KC^{-1}_1\delta' m)
$$
to $\omega(Y,Y')$, which vanishes by (\ref{SC}).
\end{proof}

\subsection*{Proof of Proposition 5}

The manipulations are slightly more transparent in the classical 
variational notation, although it is straightforward to translate this into 
the language of differential forms.

We shall evaluate $\sigma (Y,Y')$ in (\ref{sigY}) by 
putting $\Omega=\delta y y^{-1}$, $\Omega'=\delta' y y^{-1}$. We shall then
shrink the contour to the origin and use the asymptotic 
behaviour of the $y_i$s. 

For each $i$, choose $z_i\in {\cal S}_i\cap {\cal S}_{i+1}$ on the contour with $z_{2r+1}=z_1$, and 
define $\log z$ by making a cut along the ray through the origin and $z_1$.
On each sector, 
$$ 
\Omega=\Omega_{i}+y_{i}\delta C^{\str}_{i}\, 
C_{i}^{-1}y_{i}^{-1}
\, ,
$$
where $\Omega_{i}=\delta y_{i}\, y_{i}^{-1}$.
Moreover, 
$$
\nabla\bigl(y^{\str}_{i}\delta C^{\str}_{i}C_{i}^{-1}y_{i}^{-1}\bigr)=0\, , 
$$ 
since $C_{i}$ is independent of $z$.
Therefore, in the notation of (\ref{sigY}),
\be
\frac{1}{2\pi\ri}\oint \tr(\Omega\nabla\Omega') 
=\frac{1}{2\pi\ri}\sum_1^{2r}\int_{z_i}^{z_{i+1}}
\tr\bigl(\Omega^{\str}_{i}
\nabla \Omega'_{i} \bigr) 
+\frac{1}{2\pi\ri}\sum_1^{2r}\Bigl(x_i(z_{i+1})-x_{i}(z_i)\Bigr)
\label{sigma}\ee
where $z_{i}$ is some point on the contour in ${\cal S}_{i}\cap 
{\cal S}_{i+1}$,
$x_i= \tr(\delta C^{\str}_iC_i^{-1} y_i^{-1}\,\delta' y^{\str}_i)$,
 and the integrals are along segments  of the contour. 
However,
\be
x_{i+1}-x_i=\tr\Bigl(
\delta C^{\str}_{i}\, \delta C_{i}^{-1}\delta' S^{\str}_iS_i^{-1}-
S_{i}^{-1}\, \delta S_{i}y_{i+1}^{-1}\, \delta' y_{i+1}\Bigr)\, .
\label{xx}\ee
This follows from the two relations
$y_i=y_{i+1}S_i^{-1}$ and  $C_i=S_iC_{i+1}$, which imply that
$$
\tr(\delta C_i C_i^{-1}y_i^{-1}\delta' y_i)
=\tr\Bigl(\delta C_i C^{-1}_{i}S_iy^{-1}_{i+1}(\delta' y_{i+1})S_i^{-1}
-\delta C_iC_i^{-1}\delta' S_i S_i^{-1}\Bigr)
$$
and
$$
S_i^{-1}\delta C^{\str}_i C_i^{-1}S_i=\delta C^{\str}_{i+1}C_{i+1}^{-1}
-S_i^{-1}\delta S^{\str}_i\, .
$$

As $z\to 0$ in ${\cal S}_{i}\cap {\cal S}_{i+1}$, the second term on the 
right-hand side of (\ref{xx}) goes to zero by (\ref{Stokeasym}). Moreover, 
$$
x_{2r+1}-x_1=
4\pi^2\tr(\delta m\delta' m)
+2\pi\ri\tr(
\delta C_1C_1^{-1}\delta' m)
-2\pi\ri(\delta m y_1^{-1}\delta y_1)\, .
$$

To deal with
the first term in (\ref{sigma}), we note that
$$
\Omega_i=\Theta +g\delta m g^{-1} \log z +O(z^N)
$$
as $z\to 0$ in ${\cal S}_i$
for some large $N$ (depending on the truncation of the formal power series).
Therefore in ${\cal S}_i$
\begin{eqnarray*}
\lefteqn{\tr(\Omega_i \nabla \Omega'_i)=}\\
&&\tr(\Theta\nabla\Theta') + \tr\left[
\left(g^{-1}\delta g +\frac{\delta t}{z^r}\right)\delta' m
-\left(g^{-1}\delta' g +\frac{\delta' t}{z^r}\right)\delta m
\right]\frac{\rd z}{z} \\
&& {} +\p_z\Bigl(\tr\bigl(\delta'm(g^{-1}\delta g +z^{-r}\delta t)\bigr)\log 
z\Bigr)\, \rd z+\tr(\delta m\delta' m)\log z\frac {\rd z}{z}+O(N')\, ,
\end{eqnarray*}
for some large $N'$. We therefore have
\begin{eqnarray*}
\frac{1}{2\pi\ri}\sum_1^{2r}\int_{z_i}^{z_{i+1}}\tr(\Omega_i\nabla 
\Omega'_i)&=&
\frac{1}{2\pi\ri}\oint\tr(\Theta\nabla \Theta') 
+\tr\bigl(\gamma\delta' m -\gamma'\delta m\bigr)\\
&&\quad {}+\tr (\delta' m y_1^{-1}\delta y_1)_{z_1}+
\pi \ri\tr (\delta m\delta' m) 
+\epsilon
\end{eqnarray*}
where $\epsilon \to 0$ as the contour is shrunk.
The proposition follows by putting the two terms together, by 
shrinking the contour towards $z=0$, and by using the definition (\ref{omdef}).

\section*{References}

\setlength{\parindent}{0in}
\setlength{\parskip}{0.10in}
Ablowitz, M. J., and Clarkson, P. A. (1991). {\em Solitons, nonlinear 
evolution equations and inverse scattering.} LMS Lecture Notes, {\bf 149}. 
Cambridge University Press, Cambridge.

Boalch, P. (2000). {\em Symplectic manifolds and isomonodromic deformations}. 
(Pre\-print, SISSA, Trieste).

Dubrovin, B. (1999). Painlev\'e transcendents in two-dimensional 
topological field theory. In {\em The Painlev\'e property, one century later}.
Ed. R. Conte. CRM Series in Mathematics. Springer, New York.

Farkas, H. M., and Kra, I (1991). {\em Riemann surfaces}. 
Springer, New York.

Hitchin, N. J. (1995a). Twistor spaces, Einstein metrics and isomonodromic 
deformations.  {\em J.\ Diff.\ Geom.} {\bf 42}, 30--112.

Hitchin, N. J. (1995b). Poncelet 
polygons and the Painlev\'e equations.  In: {\em Geometry and analysis, Bombay 
1992}, 151--185. Tata Institute, Bombay.

Hitchin, N. J. (1996). A new family of Einstein metrics. In: {\em Manifolds 
and geometry, Pisa 1993}, 190--222.  Sympos.\ Math., XXXVI. Cambridge 
University Press, Cambridge.

Hitchin, N. J. (1997) Geometrical aspects of Schlesinger's equation. {\em J. 
Geom.\ Phys.}, {\bf 23}, 287--300.

Jimbo, M., Miwa, T., and Ueno, K. (1981a). Monodromy preserving 
deformations of 
linear ordinary differential equations with rational coefficients. I. General 
theory and $\tau$-functions. {\em Physica} {\bf 2D}, 306--52.

Jimbo, M., and Miwa, T. (1981b). Monodromy preserving deformations of 
linear ordinary differential equations with rational coefficients. II. 
{\em Physica} {\bf 2D}, 407--48.

Jimbo, M., and Miwa, T. (1981c). Monodromy preserving deformations of 
linear ordinary differential equations with rational coefficients. III. 
{\em Physica} {\bf 4D}, 26--46.

Kodaira, K. (1961). A theorem of completeness of characteristic systems for 
analytic families of compact submanifolds of complex manifolds. {\em Ann.\ 
Math.}, {\bf 75}, 146--62.

Malgrange, B. (1982). Sur les d\'eformations isomonodromiques. {\em 
S\'eminaire de l'Ecole Norm.\ Sup.}, IV, 401--26.

Mason, L. J., and Woodhouse, N. M. J. (1996). {\em Integrability, 
self-duality, and twistor theory}, Oxford.

Pressley, A.,  and Segal, G. (1986). {\em Loop groups}. Oxford University 
Press, Oxford.

Sato, M., Miwa, T., and Jimbo, M. (1978). Holonomic quantum fields. I. 
{\em Publ.\ RIMS}, Kyoto, {\bf 14}, 223-67.

Sato, M., Miwa, T., and Jimbo, M. (1979). Holonomic quantum fields. II--IV. 
{\em 
Publ.\ RIMS}, Kyoto,  {\bf 15}, 201-67, 577-629, 871--972.

Sato, M., Miwa, T., and Jimbo, M. (1980). Holonomic quantum fields. V. {\em 
Publ.\ RIMS}, Kyoto, {\bf 16}, 531--84.

Souriau, J.-M.\ (1970). {\em Structures des syst\`emes dynamique}. Dunod, 
Paris.

Woodhouse, N. M. J. (1992). {\em Geometric quantization}. 2nd edition. Oxford 
University Press, Oxford.

\end{document}